\documentclass{optica-article}
\usepackage{makecell}

\journal{opticajournal} 
\articletype{Research Article}

\usepackage{lineno}

\usepackage{textcomp} 

\begin{document}

\title{A Versatile Optical Frontend for Multicolor Fluorescence Imaging with Miniaturized Lensless Sensors}

\author{Lukas Harris\authormark{1,2,†}, Micah Roschelle\authormark{1,2,†,*}, Jack Bartley \authormark{2}, and Mekhail Anwar\authormark{1,2,*}}

\address{\authormark{1}Department of Electrical Engineering and Computer Sciences, University of California, Berkeley, California 94720, USA}
\address{\authormark{2}Department of Radiation Oncology, University of California, San Francisco, California 94158, USA}
\address{\authormark{†}These authors contributed equally to this work.}
\email{\authormark{*}micah.roschelle@berkeley.edu}
\email{\authormark{*}mekhail.anwar@ucsf.edu} 


\begin{abstract*} 
Lensless imaging enables exceptionally compact fluorescence sensors, advancing applications in \textit{in vivo} imaging and low-cost, point-of-care diagnostics. These sensors require a filter to block the excitation light while passing fluorescent emissions. However, conventional thin-film interference filters are sensitive to angle of incidence (AOI), complicating their use in lensless systems. Here we thoroughly analyze and optimize a technique using a fiber optic plate (FOP) to absorb off-axis light that would bleed through the interference filter while improving image resolution. Through simulations, we show that the numerical aperture (NA) of the FOP drives inherent design tradeoffs: collection efficiency improves rapidly with a higher NA, but at the cost of resolution, increased device thickness, and fluorescence excitation efficiency. To illustrate this, we optimize two optical frontends with full-width at half maximums (FWHMs) of 8.3° and 45.7°. Implementing these designs, we show that angle-insensitivity requires filters on both sides of the FOP, due to scattering. In imaging experiments, the 520-$\mu$m-thick high-NA design is 59× more sensitive to fluorescence while only degrading resolution by 3.2×. Alternatively, the low-NA design is capable of three-color fluorescence imaging with 110-$\mu$m resolution at a 1-mm working distance. Overall, we demonstrate a versatile optical frontend that is adaptable to a range of applications using different fluorophores, illumination configurations, and lensless imaging techniques. 
\end{abstract*}

\section{Introduction}

Lensless imaging enables exceptionally compact sensors with a relatively high resolution over a large field of view \cite{greenbaum_imaging_2012,boominathan_recent_2022}. When paired with multiplexed fluorescence imaging, these sensors are capable of high-contrast detection of multiple biomarkers, advancing an array of biomedical applications. For example, implantable or minimally invasive imagers allow for real-time recording of neural activity \cite{wu_bio-flatscope_2021, pollmann_subdural_2024}, monitoring of disease progression and therapeutic response \cite{rabbani_towards_2024,roschelle_wireless_2024, zhu_ingestible_2023}, and intraoperative imaging of cancer during surgery \cite{papageorgiou_chip-scale_2020,roschelle_multicolor_2024}. Moreover, compact and low-cost fluorescence sensors could facilitate high-throughput, point-of-care diagnostics \cite{xiahou_-chip_2025,aghlmand_65-nm_2023}.

A critical component of these sensors is an optical filter, which selectively blocks excitation light, while passing the fluorescent emissions. For organic fluorophores, the emissions can be up to 6 orders of magnitude weaker than the excitation flux and are red-shifted as little as 10 nm from the excitation wavelength. Thus, an ideal filter should have an optical density (OD) around 6 at the excitation wavelength and a rapid roll-off between rejection and pass bands for efficient fluorescence imaging. In addition, the filter should be versatile, accommodating multiplexed imaging across the visible and near-infrared spectrum. Conventional lensed systems use thin-film interference filters to meet these specifications. However, these filters are angle-sensitive; as the angle of incidence (AOI) increases, the filter passband blue-shifts towards the excitation wavelength, leading to excitation bleed-through. This property is problematic for lensless systems, where the AOI is not precisely controlled by lenses \cite{dandin_optical_2007}. 

Therefore, most prior lensless fluorescence imaging work has focused on alternative filter designs. For example, absorption filters composed of a dye \cite{yildirim_implementation_2017} or semiconductor \cite{papageorgiou_angle-insensitive_2018} are inherently angle-insensitive but suffer from gradual roll-offs and autofluorescence. Hybrid designs combining absorption and interference filters can eliminate out-of-band autofluorescence and improve the rejection-band characteristics \cite{moazeni_mechanically_2021, sasagawa_highly_2018,marin-lizarraga_simultaneous_2025}. However, absorption spectra are material-dependent and, thus, are not easily adapted to different fluorophores or for multiplexed imaging \cite{kulmala_lensless_2022, taal_toward_2022}. Another alternative, plasmonic filters \cite{aghlmand_65-nm_2023,hong_nano-plasmonics_2018}, can be integrated on-chip in the metal layers, but generally exhibit worse roll-off and rejection than absorption or interference filters. Instead of a filter, total internal reflection (TIR) optics can be used to confine the excitation light at the sensor surface \cite{shin_miniaturized_2023}, but this approach is susceptible to scattering which is unavoidable with thick biological samples. Another filter-less technique is time-gated imaging \cite{choi_fully_2020, najafiaghdam_optics-free_2022}, where the laser is pulsed prior to imaging. However, this approach is challenging with organic fluorophores which have a fluorescence lifetime on the order of 1 ns \cite{berezin_fluorescence_2010}. Thus, a high-performance and angle-insensitive filter design that is adaptable to different fluorophores and for multiplexed imaging remains an elusive design goal.

\begin{figure}[tbp]
\centering\includegraphics[]{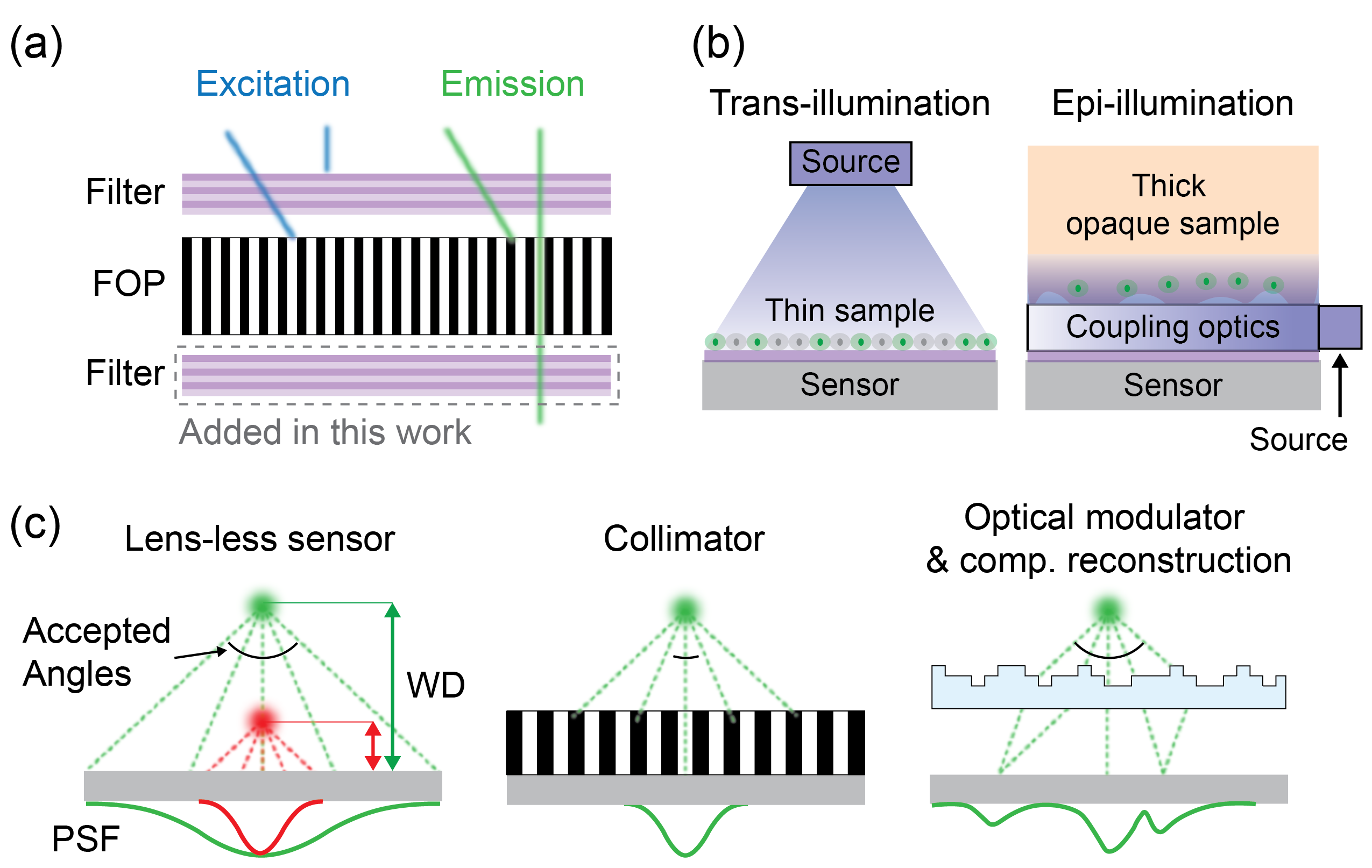}
\caption{\textbf{Proposed optical frontend design for lensless fluorescence imaging. } (\textbf{a}) Optical frontend design consisting of a fiber optic plate (FOP) with interference filter coatings on one or both sides. (\textbf{b}) When imaging thin samples, trans-illumination is possible. Otherwise epi-illumination is required. (\textbf{c}) The resolution, or point spread function (PSF), can be improved by reducing the working distance (WD), adding a collimator, or using a modulating optic with computational reconstruction. }
\label{concept}
\end{figure}

To address this need, we improve on a technique originally presented in \cite{roschelle_multicolor_2024} which uses a low-numerical-aperture (NA) fiber optic plate (FOP) to compensate for the angle-dependence of interference filters (Fig. \ref{concept}(a)). In this approach, the FOP acts as an angle filter passing light near normal incidence for imaging, while absorbing off-axis light that would otherwise bleed through the interference filter. The FOP also improves resolution by eliminating divergent rays that contribute to blur. In our prior work \cite{roschelle_multicolor_2024,roschelle_wireless_2024}, we demonstrate a 520-$\mu$m-thick optical frontend optimized for \textit{in vivo} fluorescence imaging, achieving OD 6 excitation rejection as well as three-color fluorescence imaging with 110-$\mu$m resolution. However, this design is still sensitive to AOIs less than 5° and uses a low-NA FOP with a full-width at half-maximum (FWHM) of less than 10° to improve resolution, but at a severe cost to the total collected fluorescence emissions.

Here, we undertake a comprehensive design-space exploration of this technique, revealing how it can be (1) modified to block excitation light across all AOIs allowing for different illumination configurations (Fig. \ref{concept}(b)) and (2) adapted to accept a wider range of AOIs to improve light collection efficiency and accommodate computational imaging techniques (Fig. \ref{concept}(c)). The AOI of the excitation light depends on the illumination configuration. For \textit{in vitro} imaging of thin samples, the sample can be trans-illuminated at normal incidence through its backside. However, imaging thick tissue samples (as in \textit{in vivo} applications) requires epi-illumination with the light typically introduced at oblique AOIs between the sample and the sensor. Epi-illumination involves coupling optics, such as a light-guide plate \cite{shin_miniaturized_2023,sasagawa_front-light_2023}, to guide light from edge-mounted sources, increasing the working distance and leading to reduced resolution (Fig. \ref{concept}(c)). However, the resolution can be improved with either a collimator like a FOP or an optical modulator, such as a phase mask \cite{adams_vivo_2022} or diffuser \cite{antipa_diffusercam_2018,kuo_-chip_2020}, and iterative reconstruction algorithms. While a collimator improves resolution by eliminating light at divergent angles, computational imaging approaches exploit larger AOIs to generate a sparse point spread function (PSF) with a large spatial extent. 

In this work, we analyze how these system-level specifications motivate the choice of specific filter and FOP design parameters. For the FOP, the driving design parameter is the NA: a higher NA expands the range of accepted angles, improving collection efficiency, but at a cost to resolution. While a high NA is desired for single-pixel sensors or computational imaging approaches, a low-NA design is preferred for imagers that rely on the FOP for resolution. Through ray-tracing simulations, we show how the thickness and fill factor of the FOP can be adjusted to accommodate FOPs with a range of NAs. We then implement and characterize two multi-bandpass optical frontend designs using FOPs at the extremes of this design space with NAs of 0.15 and 0.43 corresponding to FWHMs of 8.3° and 45.7°, respectively. Both designs are optimized for minimum thickness, while still providing at least OD 6 of excitation rejection. In implementing these designs, we find that coating the FOP with a single interference filter, as done previously, results in a design that is sensitive to specific AOIs due to scattering. Thus, the relative order of the filter and the FOP must be selected according to the illumination configuration. To address this issue, we design a new optical frontend with filters coated on both sides of the FOP to achieve angle-insensitive excitation rejection. Finally, we illustrate the tradeoffs in resolution and collection efficiency for both optical frontend designs by imaging a USAF target, fluorophore dilution series, and three-color fluorescent beads.

\section{Optical frontend design}

\subsection{Interference filter characterization}

Thin-film interference filters are constructed by layering materials of alternating refractive indices with their thickness precisely tuned to create wavelength-specific interference at each interface. Thus, when viewed at an angle, the effective layer thicknesses change, causing the filter spectra to blue-shift. For a simple single-layer interference filter (or Fabry-Pérot cavity) with center wavelength at normal incidence, $\lambda_0$, the shift in the center wavelength at increasing AOIs, $\theta$, can be modeled as:
\begin{equation}
\lambda(\theta)=\lambda_0\sqrt{1 -(\frac{\sin\theta}{n})^2} 
\end{equation}
where $n$ is the refractive index of the cavity. 

\begin{figure}[tbp]
\centering\includegraphics[]{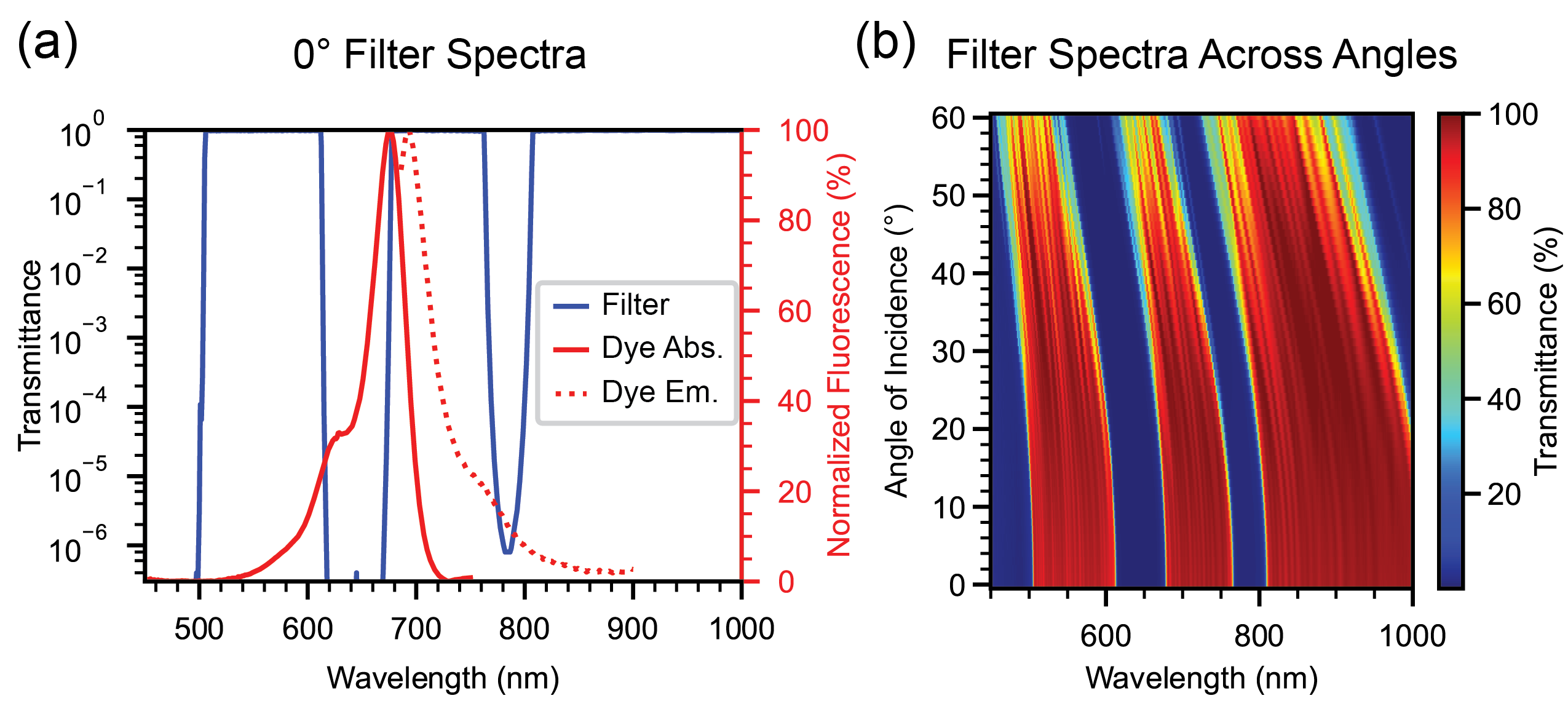}
\caption{\textbf{Interference filter characterization. }(\textbf{a}) Transmittance spectra of the interference filter at normal incidence with the absorption and emission spectra of Licor IRDye 680LT. (\textbf{b}) Measured filter transmittance spectra across incident angles. }
\label{int filter}
\end{figure}

In this work, we use a commercially available thin-film interference filter coating (ZET488/647\hspace{0pt}/780+800lpm Chroma Technology Corp.) with its spectrum shown in Fig. \ref{int filter}(a). (Data is reproduced with permission from Chroma Technology Corp.) The filter is 19.4 $\mu$m thick and has three pass-bands, covering 505 - 612 nm, 677 - 763 nm, and 807 nm long-pass. For reference, the absorption and fluorescence emission spectra of IRDye 680LT ($\lambda_{EX}$=677 nm, $\lambda_{EM}$=694 nm)–which we use in our imaging experiments–is overlaid on the plot. At normal incidence, the filter has more than OD 6 rejection in the stop bands, steep roll-offs, and greater than 95\% transmittance in the passbands (see Fig. \ref{lin filter spectra}), allowing for efficient fluorescence imaging.

Fig. \ref{int filter}(b) shows the measured blue-shift of the filter spectra with increasing AOIs. The measurement setup is shown Fig. \ref{spectra setup}. Following Eq. (1), the blue-shift is minor for small AOIs, but accelerates for larger AOIs. While choosing an excitation wavelength further from the band-edge ensures blocking for a wider range of angles, the fluorophore is now excited further from its absorption peak, reducing the fluorescent flux. For instance, a 635 nm excitation source bleeds through the filter with more than 10\% transmittance at 42° while for a 660 nm source this occurs at 24°. On the other hand, IRDye 680LT has an extinction coefficient of 33\% of its peak value at 635 nm and 65\% of the peak at 660 nm. 

\subsection{Compensating for the filter and improving resolution with a collimator}

We can compensate for the angle-dependent spectra of the interference filter by adding an angle filter, or collimator, which blocks the oblique excitation light that would bleed through the interference filter while still passing fluorescence emissions close to normal incidence for imaging. The collimator will also block divergent fluorescence emission that contribute to blur, improving resolution, albeit at a cost to collection efficiency. 

To understand this inherent tradeoff between resolution and collection efficiency, we derive expressions for a lensless imager sensor using a collimator with an angular transmittance of $T(\theta)$ In general, the resolution of an imaging system is determined by the point spread function (PSF), which, in this case, can be derived as $PSF(\theta)=T(\theta) \cdot \cos^3 \theta$ (Supplementary Section 1). Typically, the resolution of a given PSF is quantified by its full-width at half maximum, $\theta_{FWHM}$. Thus, for a sample at a distance $l$ from the sensor, the spatial resolution is $FWHM_{X,Y}=2l \cdot \tan \frac{\theta_{FWHM}}{2}$. On the other hand, the collection efficiency, $\eta_C$, is defined as the percentage of total emitted light that is collected by the sensor and can be derived as,
\begin{equation}
 \eta_C = \frac{1}{2}\int_0^{\pi/2} T(\theta) \sin(\theta) d\theta
\end{equation}
(Supplementary Section 2). 

It is instructive to examine these expressions for an ideal angle filter with a rectangular $T(\theta)$ defined by a small cutoff angle, $\theta_C$ (Fig. \ref{fop sims}(a)). Using small-angle approximations, the resolution is $FWHM_{X,Y} \approx 2l \cdot \tan\theta_C \approx 2l \cdot \theta_C$ and the collection efficiency is $\eta_C = \sin^2 \frac{\theta_C}{2} \approx \frac{\theta_C^2}{4}$ (Supplementary Section 2). Therefore, decreasing $\theta_C$ linearly improves resolution, but comes at a much larger (approximately quadratic) cost to collection efficiency. Alternatively, the resolution can be improved by reducing $l$ at no cost to collection efficiency. However, in many devices, epi-illumination optics and the filter thickness will enforce a non-zero sample-to-sensor distance.

In practice, this angle filter can be implemented with a thin and planar form factor suitable for lensless imaging as a parallel-hole (or Söller) collimator \cite{balsam_lensless_2011}, which consists of an array of transparent holes in an absorptive material. However, a sharper cutoff can be achieved with a FOP, which takes advantage of total-internal reflection (TIR).

\subsection{Design and simulation of fiber optic plates}

\begin{figure}[tbp]
\centering\includegraphics[]{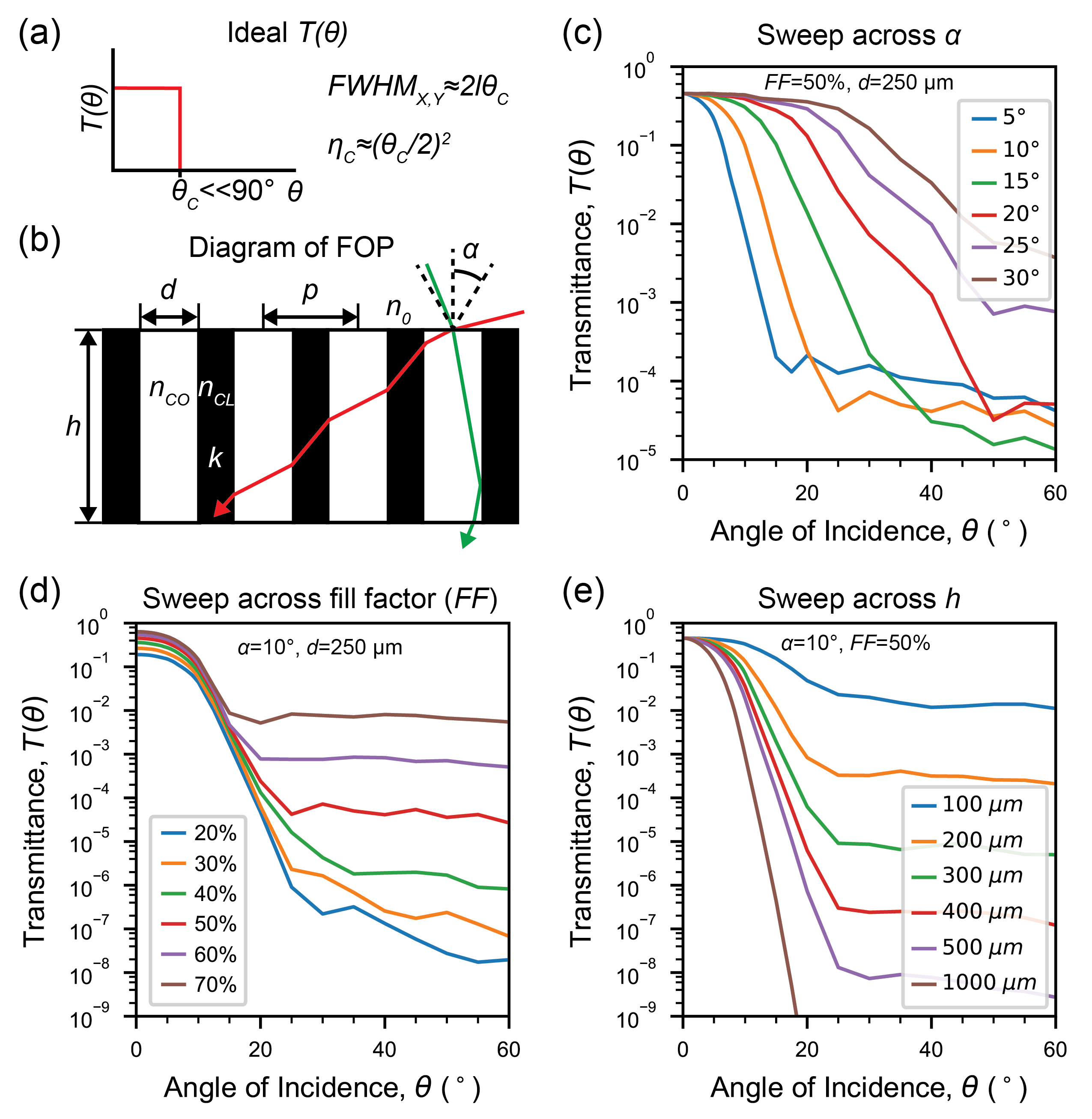}
\caption{\textbf{FOP Simulations.} (\textbf{a}) Ideal angle filter response, $T(\theta)$, and the resulting resolution ($FWHM_{X,Y}$) and collection efficiency ($\eta_{C}$). (\textbf{b}) Diagram of a FOP showing key design parameters: acceptance angle, $\alpha$; fiber diameter, $d$; fiber pitch, $p$; and thickness, $h$. (c-e) Simulated transmittance across AOIs for different (\textbf{c}) fiber acceptance angles ($\alpha$), (\textbf{d}) fill factors ($FF$), and (e) thicknesses ($h$). }
\label{fop sims}
\end{figure}

As shown in Fig. \ref{fop sims}(a), the FOPs used in this work consist of a hexagonal array of optical fibers with a transparent core and black, absorptive cladding. For light at normal incidence, the transmittance depends on the fill factor, $FF$, of the FOP, which is defined as the percentage of the total surface area that is transparent. $FF$ is a function of the fiber core diameter, $d$, and the fiber pitch, $p$. For AOIs within the acceptance angle of the fibers, $\alpha$, light is guided through the FOP by TIR, while for AOIs beyond $\alpha$, light passes through the absorptive cladding. $\alpha$, is determined by the difference between the lower index of refraction of the cladding, $n_{CL}$, and the higher refractive index of the core, $n_{CO}$, according to
\begin{equation}
\alpha = \sin^{-1}\left(\frac{1}{n_0} \sqrt{n_{CO}^2 - n_{CL}^2}\right)
\end{equation}
where $n_0$ is the refractive index of the incident medium. Often, the range of accepted angles is reported as the numerical aperture, $NA=\sin \alpha$. 

Light that does not experience TIR has transmittance, $T$, through the FOP according to the Beer-Lambert law: $T=e^{-k \cdot OP}$, where $k$ is the absorption coefficient of the cladding and $OP$ is the optical path-length traversed through the cladding. Here, $k$ is assumed to be maximized within the limits of the manufacturing process. Therefore, stronger rejection can be achieved by lengthening $OP$ either by increasing the plate thickness, $h$, or by decreasing $FF$ of the FOP. 

Through ray-tracing simulations with TracePro (Lambda Research Organization), we illustrate how the FOP parameters ($\alpha$, $FF$, and $h$) can be adjusted to design an angular transmittance function, $T(\theta)$, that effectively compensates for the angular sensitivity of the filter while considering tradeoffs between resolution and collection efficiency. In our simulations, we measure $T(\theta)$ of the FOP by sweeping the incident angle of a collimated 660 nm pencil beam. We perform parameter sweeps across $\alpha$, $FF$, and $h$ for FOPs with the following default parameters: $h=250 \text{ $\mu$m}$, $d=20 \text{ $\mu$m}$, $p=27 \text{ $\mu$m}$, $n_{CL}=1.56$, and $n_{CO}=1.57$, resulting in $FF=50\%$, $\alpha=10^{\circ}$, and $NA=0.17$. $k$ is derived from the measured transmittance through the cladding material used in the high-NA FOP from our physical experiments.

Fig. \ref{fop sims}(b) shows how the choice of $\alpha$ impacts $T(\theta)$ of the FOP. At angles beyond $\alpha$, $T(\theta)$ rapidly decreases until absorption saturates due to refraction at the air-FOP interface. The angle at which $T(\theta)$ reaches this saturation point increases with larger $\alpha$. To effectively compensate for the filter, the FOP should reach its maximum absorption at the angle at which the excitation wavelength completely transmits through the interference filter. Given that wavelengths closer to the band-edge bleed through the filter at smaller angles, a lower $\alpha$ enables excitation closer to the filter band-edge for more efficient fluorescence imaging. As described in the preceding section, $\alpha$ also drives the tradeoff between resolution and collection efficiency. Fig. \ref{lin sim results}(a) shows $T(\theta)$ on a linear scale with the calculated resolution and collection efficiency for each $\alpha$  in Fig. 
\ref{lin sim results}(b). As expected, increasing  dramatically improves collection efficiency while reducing resolution.

$FF$ and $h$ control the absorption of the FOP at AOIs beyond $\alpha$. Since Beer's law shows an exponential dependence on optical path-length, a linear increase in either of these parameters results in an exponential increase in absorption. This trend is observed in the simulations of $T(\theta)$ while sweeping $FF$ and $h$ across linear ranges as shown in Figs. \ref{fop sims}(c) and \ref{fop sims}(d), respectively. Decreasing $FF$ to improve off-axis absorption comes at the cost of reduced transmittance at normal incidence as is clearly seen in the linear-scale plot in Fig. \ref{lin sim results}(c). Alternatively, increasing $h$ has no effect on the normal incidence transmittance, but results in a bulkier device. As shown in Fig. \ref{lin sim results}(d), increasing $h$ also slightly reduces the FWHM of the FOP. However, this effect can be compensated for by slightly increasing $\alpha$.

In conclusion, the acceptance angle (or NA) of the FOP should be selected depending on (1) the angular dependence of the filter which is determined by the proximity of the excitation wavelength to the filter band-edge and (2) the desired resolution and collection efficiency of the sensor. Generally, a larger $\alpha$ improves collection efficiency, but comes at a cost to both image resolution and the excitation efficiency of the fluorophore. Once a specific $\alpha$ is selected, the fill factor and plate thickness should be adjusted such that the FOP achieves enough absorption at oblique AOIs to effectively compensate for excitation bleed-through from the filter. Practically, greater rejection comes at the cost of reduced normal incidence transmittance (when changing $FF$) and increased device thickness (when modifying $h$).

The final design consideration for the FOP is the selection of the fiber pitch. As described in Supplementary Section 3, to achieve pixel-limited resolution, the fiber pitch must be less than the pixel pitch. Furthermore, to minimize the pixel-level intensity modulation due to imperfect alignment between the pixels and fibers, we show that the fiber pitch should be less than the pixel pitch by at least a factor of 2 (see Fig. \ref{fiber sizing}).

Finally, it is important to note that the above ray-optics simulations do not account for scattering due to diffraction through the fiber apertures, material imperfections in the fiber cores, or the surface roughness of the FOP \cite{xia_wave_2020}. As discussed later, these effects have a significant impact on the performance of the FOP and the interference filter.

\section{Optical frontend implementation}

\begin{figure}[tbp]
\centering\includegraphics[]{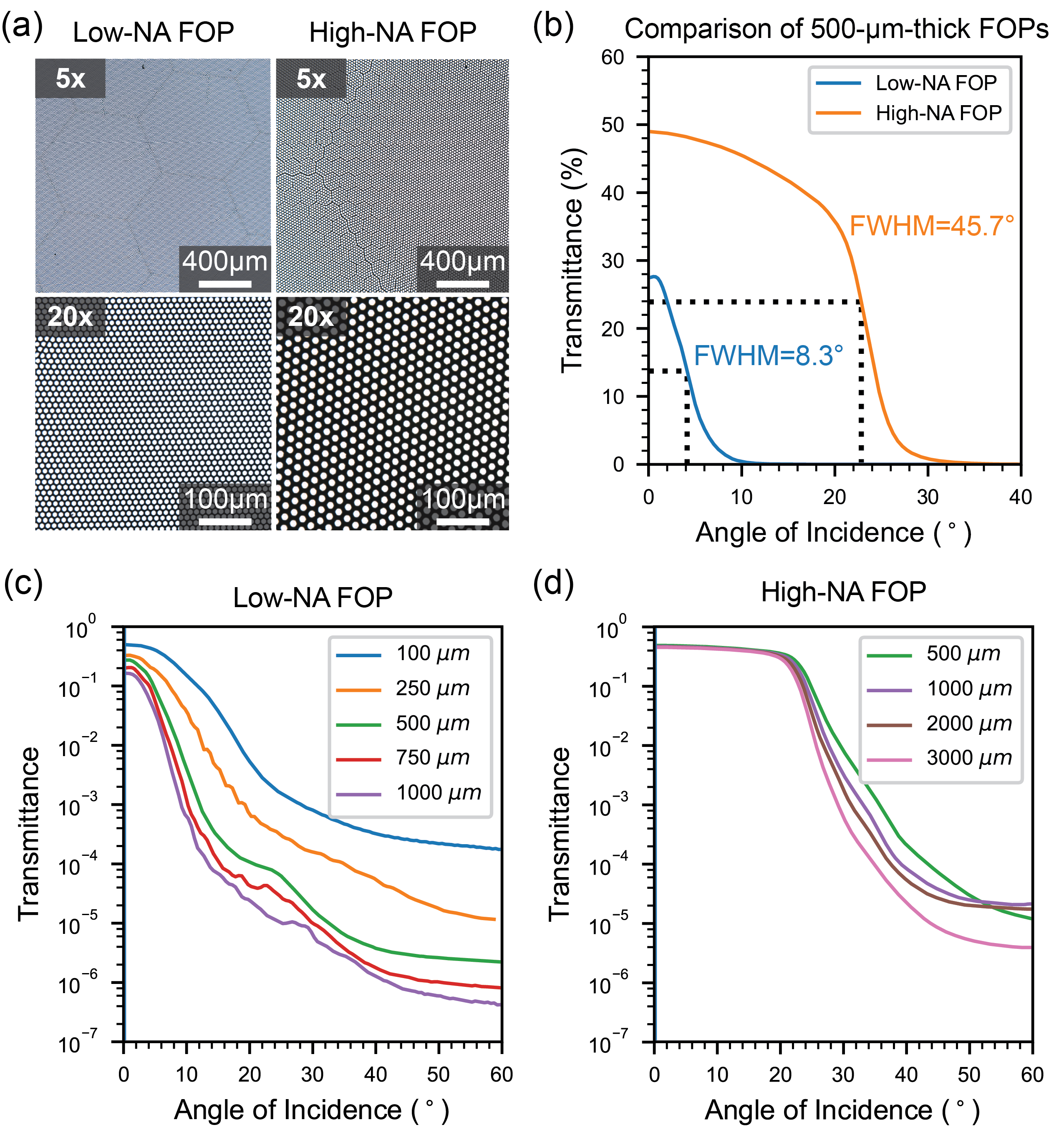}
\caption{\textbf{FOP Characterization.} (\textbf{a}) Micrographs of both FOPs with 5× and 20× magnification. (\textbf{b}) Measured transmittance across AOIs of a collimated 660 nm laser for both 500-$\mu$m-thick FOPs. (\textbf{c},\textbf{d}) Measured transmittance across AOIs for different thicknesses of the (c) low- and (d) high-NA FOPs. }
\label{fop char}
\end{figure}

\subsection{Fiber optic plate characterization}
To illustrate the tradeoffs shown in our simulations, we characterize two FOP designs with NAs at the extremes of this design space. The low-NA design (LNP121011, Shenzhen Laser LTD), which was used in our prior work \cite{roschelle_multicolor_2024,roschelle_wireless_2024}, has a NA of 0.15, a fiber pitch of 11 $\mu$m, a fiber core diameter of 9 $\mu$m, and core refractive index of 1.51. The high-NA design (J12221, Hamamatsu Corp) has a NA of 0.43, a fiber pitch of 20 $\mu$m, fiber core diameter of 14.5 $\mu$m, and a core refractive index of 1.57. Fig. \ref{fop char}(a) shows micrographs of both designs. The fiber diameter and pitch of both designs were selected from the limited options from each manufacturer, such that they were significantly less than the pixel pitch of our sensor (55 $\mu$m) and resulted in a similar fill factor for both plates.  

To determine the minimum thickness for each design that still effectively compensates for the interference filter, we measure the angular transmittance of each FOP across different thicknesses: 100, 250, 500, 750, and 1000 $\mu$m for the low-NA FOP and 500, 1000, 2000, and 3000 $\mu$m for the high-NA FOP. The resulting measurements are plotted in Fig. \ref{fop char}(b-d). These are performed by measuring the transmittance of a collimated 660 nm fiber-coupled laser (QFLD-660-10S, QPhotonics) through the FOP with an optical power meter (PM100D with S171C, Thorlabs) as the FOP is rotated between 0° and 60° on a motorized rotation stage (HDR50, Thorlabs). A diagram of the setup is shown in Fig. \ref{trans meas setup}. The optical power output of the laser is 4.23 mW. 

Fig. \ref{fop char}(b) compares the measured angular transmittance of the 500 $\mu$m-thick low- and high-NA FOPs on a linear scale. In air, the low- and high-NA FOPs exhibit FWHMs of 8.3° and 45.7°, respectively. Notably, the high-NA FOP has almost double the 0° transmittance as the low-NA FOP (27\% compared to 49\%) due to differences in the fill factor and optical transmittance of the fibers in each design. Furthermore, Fig. \ref{trans spectra}(a) shows the measured 0° transmittance across the spectra for both FOPs. While both FOPs see a reduction in transmittance at longer wavelengths, this effect is particularly pronounced for the low-NA FOP, dropping from 40\% at 450 nm to 19\% at 800 nm. 

Figs. \ref{fop char}(c) and \ref{fop char}(d) show the measured angular transmittance for each thickness of the low- and high-NA FOPs, respectively. As in our simulation results, increasing the NA of the FOP extends the angle at which the FOP reaches maximum absorption. Furthermore, thicker FOPs are more absorptive of off-axis light. However, the difference in transmittance between FOPs with different thicknesses, especially at high angles, is markedly less than the exponential relationship predicted by the Beer-Lambert law and shown in our simulations. This is likely because the transmittance at large AOIs is not limited by absorption, but rather from scattering within the FOP. At oblique AOIs, light within the FOP could scatter to an angle within the acceptance angle of the fibers such that it totally internally reflects through the FOP. This effect is likely more significant for the high-NA FOP as it has a larger acceptance angle, making it more susceptible to a wider range of scattered angles. However, it could also be partly due to differences in the design  and manufacturing of each FOP. 

Figs. \ref{lin thickness sweeps}(a) and \ref{lin thickness sweeps}(b) show the same measurements in Fig. \ref{fop char}(c,d) but on a linear scale. As predicted in the simulations, the FWHM of both FOPs decreases slightly with thickness. The 0° transmittance also reduces with thickness due to attenuation in the fiber cores, especially for the low-NA FOP. For instance, between the 100 and 1000 $\mu$m-thick low-NA FOP, the 0° transmittance reduces from 50\% to 16\%. This effect is less noticeable in the high-NA FOP due to differences in material choice and fabrication processes between the two manufacturers.

\subsection{Optical frontend characterization}

\begin{figure}[tbp]
\centering\includegraphics[]{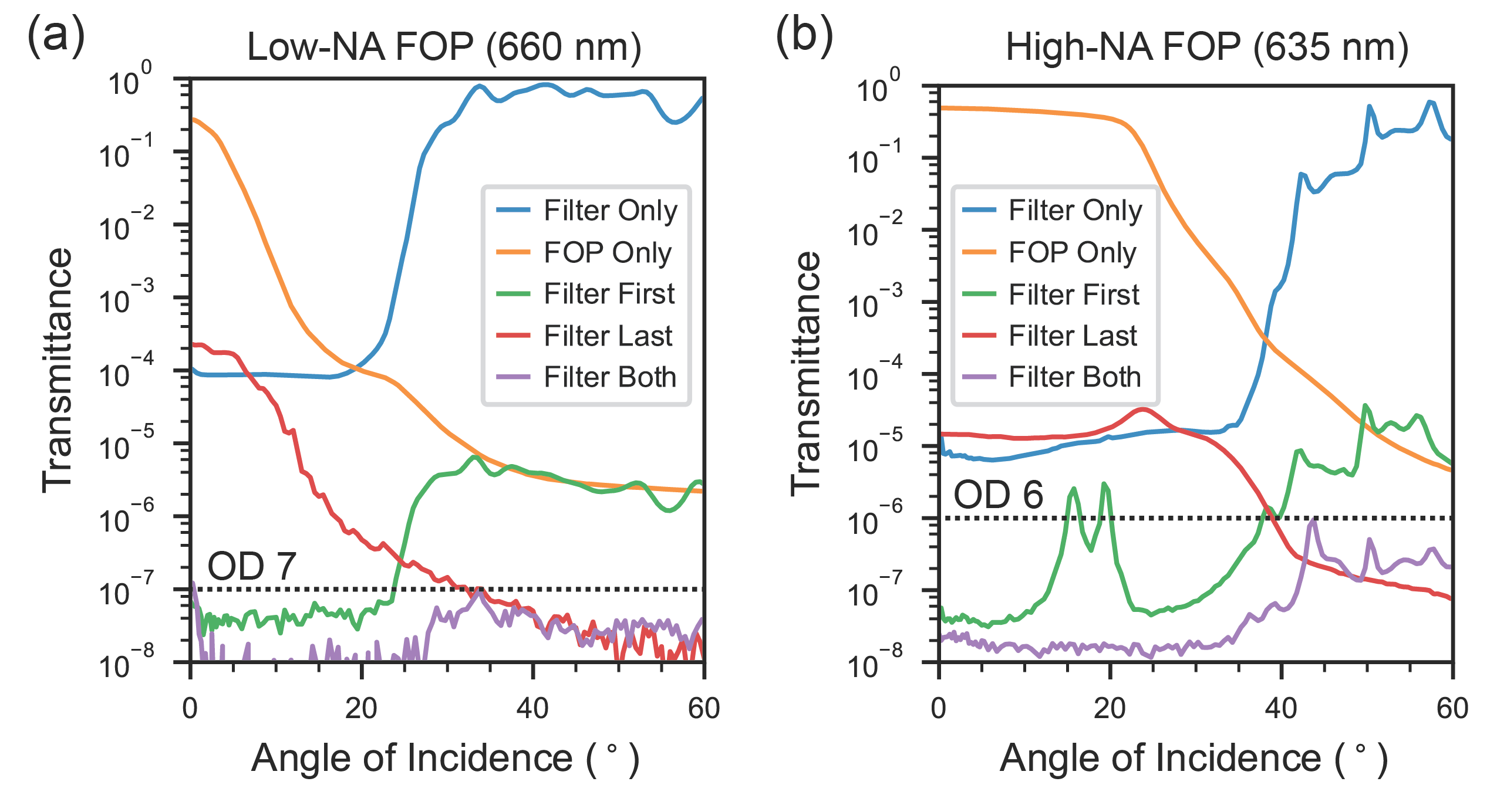}
\caption{\textbf{Optical frontend characterization.} (\textbf{a}) Measured transmittance across AOIs of a collimated 660 nm laser through the 500-$\mu$m-thick low-NA FOP, the interference filter, and different configurations of the filter and the FOP. (\textbf{b}) The same measurement for the high-NA FOP, but with a 635 nm laser.}
\label{of char}
\end{figure}

To fabricate the optical frontends, we directly coat the interference filter on either one or both sides of the FOP using a modified magnetron sputtering process through a commercial vendor (Chroma Technology Corp.). Since the deposition process exerts a mechanical force which can deform the substrate if it is too thin, the minimum FOP thickness for coating in our process is 500 $\mu$m. However, thinner designs are achieved by grinding and polishing the backside of the substrate after deposition through a commercial vendor (Mark Optics Inc), but only in the case of a single-sided coating. 

To demonstrate that both FOPs effectively compensate for the angle-sensitivity of the interference filter, we measure the angular transmittance of the excitation laser through different configurations of the filter and the FOPs. Figs. \ref{of char}(a) and \ref{of char}(b) show the measurement results for the 500-$\mu$m thick low- and high-NA FOPs, respectively. For each, we show the angular transmittance of the filter, the FOP, and three different configurations of the optical frontend (refer to the diagram in Fig. \ref{filt or}(a)): (1) with the filter first in the light path before the FOP, (2) with the filter last in the light path, and (3) with the filter on both sides of the FOP. The measurement setup is similar to that in Fig. \ref{trans meas setup} and uses either a 660 nm or 635 nm fiber-coupled laser operated at 4.23 and 15.63 mW, respectively. In addition, narrow bandpass filters are placed in the laser light path to eliminate any out-of-band emissions (ET660/20m and ZET633/20x, Chroma Technology Corp.).

As shown in Fig. \ref{of char}(a), with the low-NA FOP, the 660 nm laser starts to bleed through the filter around 20° and almost completely transmits at 30°. Near 0°, the filter only achieves OD 4 (instead of at least OD 6 as expected from Fig. \ref{int filter}(a)) due to scattering on the surface of the filter, demonstrating that even with a collimated normal incident excitation source, an interference filter alone does not provide adequate rejection. However, when the interference filter is coated on both sides of the FOP, we observe at least OD 7 across all angles. This is not the case when the filter is only coated on one side of the FOP in either of the two possible configurations. With the filter first, we see a similar rejection to the dual-sided configuration near 0°, but at larger angles, it increases to OD 5. On the other hand, the filter last configuration matches the performance of the dual-coated configuration at larger angles, but only achieves OD 3.7 at 0°. 

While the single-sided configurations may not maintain optimal performance across all AOIs, we show that with these designs the low-NA FOP can be thinned to 250 $\mu$m while still maintaining adequate excitation rejection at specific AOIs. Fig. \ref{filter configurations}(a) demonstrates that these thinner designs  still achieve OD 6 rejection as long as the AOI is less than 22° or greater than 40° if using the filter first or filter last configurations, respectively.

Fig. \ref{filter configurations}(b) shows the same measurements as Fig. \ref{of char}(a), but with the 500-$\mu$m thick high-NA FOP. However, with the higher NA, the FOP does not provide enough rejection of the 660 nm laser by the angle at which the filter pass-band shifts over it. Therefore, this design requires a shorter 635 nm excitation wavelength that is further from the filter band-edge. Fig. \ref{of char}(b) shows the measurements with the 635 nm laser (QFLD-635-20S, QPhotonics). Here, we see that the design with the filter on both sides of the FOP achieves at least OD 6 across all angles. The trends with the single-sided configurations are similar to those for the low-NA FOP.  

The single-sided filter configurations do not provide maximum rejection across all angles, due to scattering within the FOP as illustrated in Fig. \ref{filt or}(a). For the filter-first configuration, light at an oblique angle almost completely transmits through the filter, and some of the light scatters through the fibers rather than being absorbed in the cladding. Therefore, the rejection at these angles is limited to scattering in the FOP. This effect is improved with the filter-last configuration as any obliquely incident light that scatters to an angle that TIR within the fibers will be blocked by the filter. However, the filter-last configuration is susceptible to any excitation light near normal incidence that is scattered to a sufficiently large angle upon exiting the FOP such that it passes through the filter. This is not a problem in the filter-first configuration because the filter attenuates the normal incident light before it enters the FOP, so any scattered light is insignificant.

Intuitively, coating the filter on both sides of the FOP mitigates both of these scattering effects. However, there will be some loss in passband transmittance due to insertion loss of the second filter. Using a white light source (SLS201L, Thorlabs), we measure that adding a second filter slightly reduces the passband transmittance by 2\%.

\section{Imaging Results}
For our imaging experiments, we use a custom CMOS sensor designed for wireless \textit{in vivo} imaging \cite{rabbani_173_2024}, but operated with wired power and data transfer in this work. The image sensor includes a 36 × 40 array of 44 × 44 $\mu$m\textsuperscript{2} pixels with a 55 $\mu$m pitch. The optical frontend is fixed to the surface of the chip with optically clear epoxy (Sylgard 184, Electron Microscopy Sciences) and sealed on all sides with black epoxy (Supreme 3HTND-2CCM, Master Bond Inc) to minimize excitation bleed through (see Fig. \ref{imaging setup}(a)). 

\subsection{Imaging with different filter configurations}

\begin{figure}[tbp]
\centering\includegraphics[]{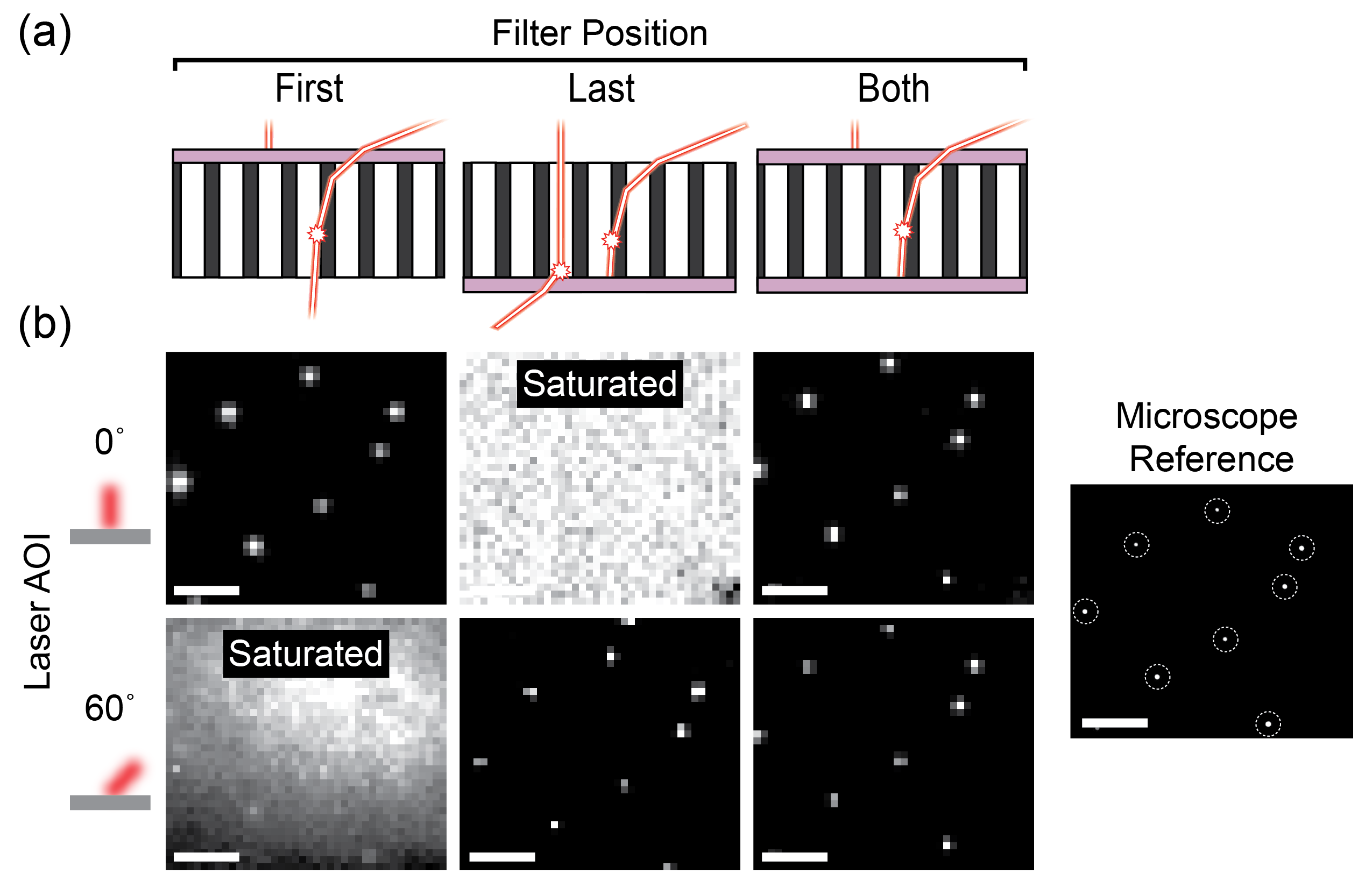}
\caption{\textbf{Imaging with different filter configurations.} (\textbf{a}) Scattering affects the performance of the optical frontend depending on the AOI of the excitation light and relative order of the FOP and filter in the light path. (\textbf{b}) Images of red fluorescent beads captured with each filter configuration in (a) and the excitation laser either at 0 or 60 degrees. The scale bar is 500 $\mu$m.}
\label{filt or}
\end{figure}

To illustrate the difference between the single- and dual-sided filter configurations, we image 15-$\mu$m-diameter red fluorescent beads ($\lambda_{EX}$=645 nm, $\lambda_{EM}$=680 nm, F8843, Thermo Fisher Scientific Inc.) using the 500-$\mu$m-thick low-NA FOP with the filter in all three configurations: filter-first, filter-last, and filter on both sides. The beads are suspended in phosphate-buffered saline (PBS) and 30 $\mu$ L of solution is pipetted into a 0.12-$\mu$m-thick chamber slide (HBW2240FL, Grace Bio-Labs) for imaging. The chamber slide is sealed with a 0.15-$\mu$m-thick coverslip and is placed with plastic film in contact with the sensor. For each filter configuration, an image is captured with the 660 nm laser trans-illuminating the sample at either 0 or 60 degrees. 
 
The imaging results are shown in Fig. \ref{filt or}(b) along with a microscope image of the beads for reference. As expected from our measurements in Fig. \ref{of char}(a), the filter-first configuration works well at normal incident excitation, but with oblique illumination, the increased excitation bleed-through saturates the sensor. The opposite is the case with the filter-last configuration, which images well with oblique illumination, but not at normal incidence. The dual-sided coated design works well in both cases. 

The following imaging experiments are performed with the optical frontends in the filter-last configuration and trans-illumination at approximately 60°.

\subsection{Collection efficiency measurements}

\begin{figure}[tbp]
\centering\includegraphics[]{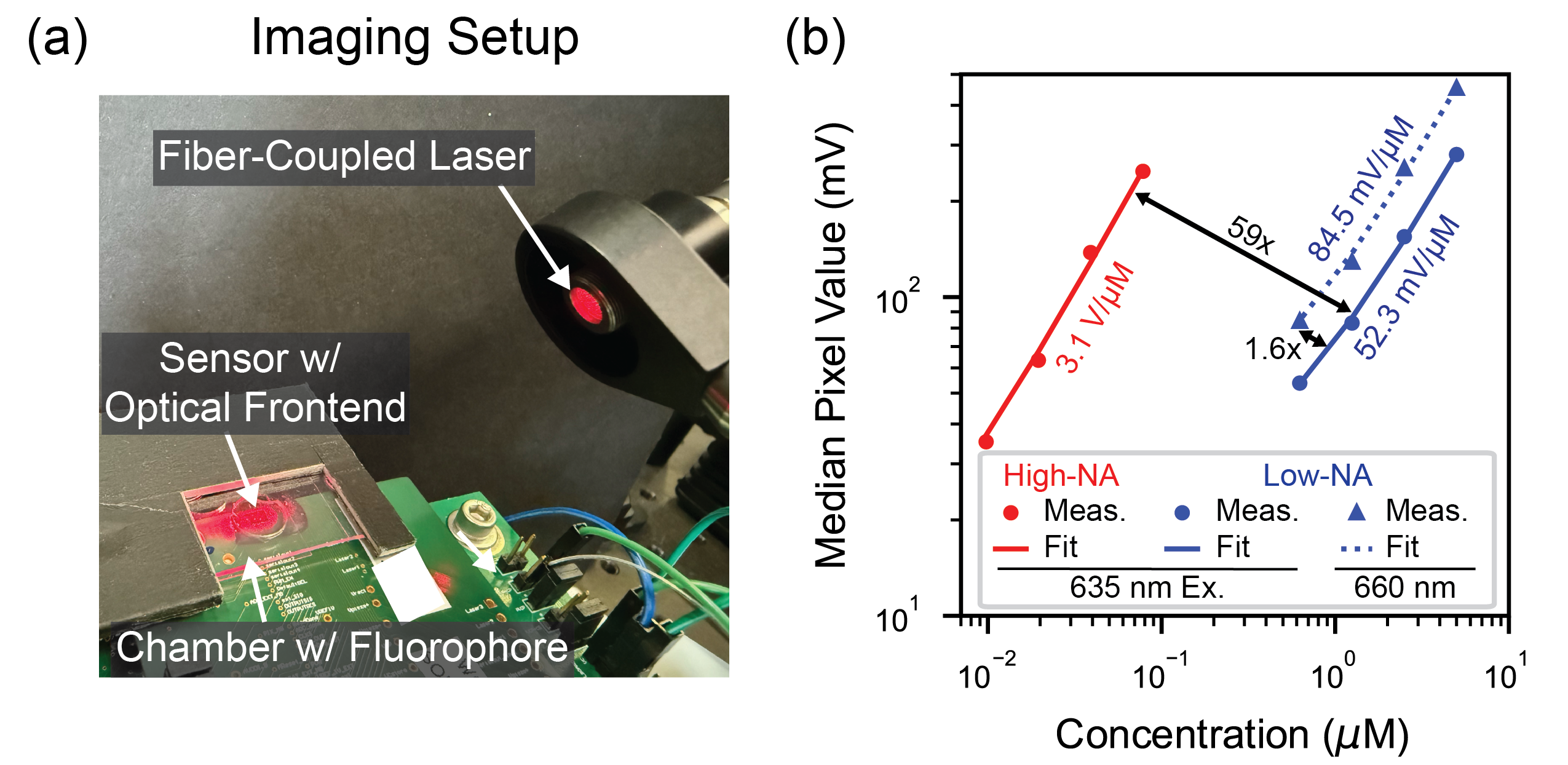}
\caption{\textbf{Collection efficiency measurements. }(\textbf{a}) Imaging setup for the experiment. (\textbf{b}) Plot of mean pixel intensity vs. fluorophore concentration across the dilution series for the low- and high-NA FOPs.}
\label{CE}
\end{figure}

We measure the collection efficiency of the low- and high-NA designs by imaging a serial dilution of IRDye 680LT NHS Ester (Licor Biosciences) in PBS. The dilutions range in concentration from 10 nM to 5 $\mu$m where each subsequent dilution is half the concentration of the preceding. 35 $\mu$ L of each solution is pipetted into 1-mm-thick chamber slides (103340, Grace Bio-Labs) that are then sealed with a 0.15-$\mu$m-thick coverslip. As shown in Fig. \ref{CE}(a), the slides are placed on sensors with either the 500-$\mu$m-thick low- or high-NA optical frontend designs for imaging and are trans-illuminated with the 635 nm laser operated at 7.12 mW/cm\textsuperscript{2}. Additionally, for the low-NA sensor, a 660-nm laser is used with the same intensity.

For each sensor, we imaged dilutions spanning the dynamic range of the sensor. Each image is  captured with an exposure time of 8 ms and is averaged 20 times to improve the signal-to-noise ratio (SNR). Fig. \ref{CE}(b) plots the median pixel intensities for each image. For each sensor, we fit a line to the data to calculate the sensitivity of the sensor in V/$\mu$m of fluorophore solution. The results show that with the high-NA FOP, the sensor collects 59× more light than with the low-NA FOP when using the 635 nm excitation laser. This ratio is reduced by 1.6× when the 660 nm laser is used with the low-NA FOP to excite the fluorophore closer to its absorption peak. The  660 nm laser is incompatible with the high-NA design. Although the higher collection efficiency of the high-NA FOP is partially due to its slightly higher 0° transmittance (see Fig. \ref{fop char}(b)), it is primarily due to the fact that it accepts a wider range of angles.

\subsection{Resolution measurements}

\begin{figure}[tbp]
\centering\includegraphics[]{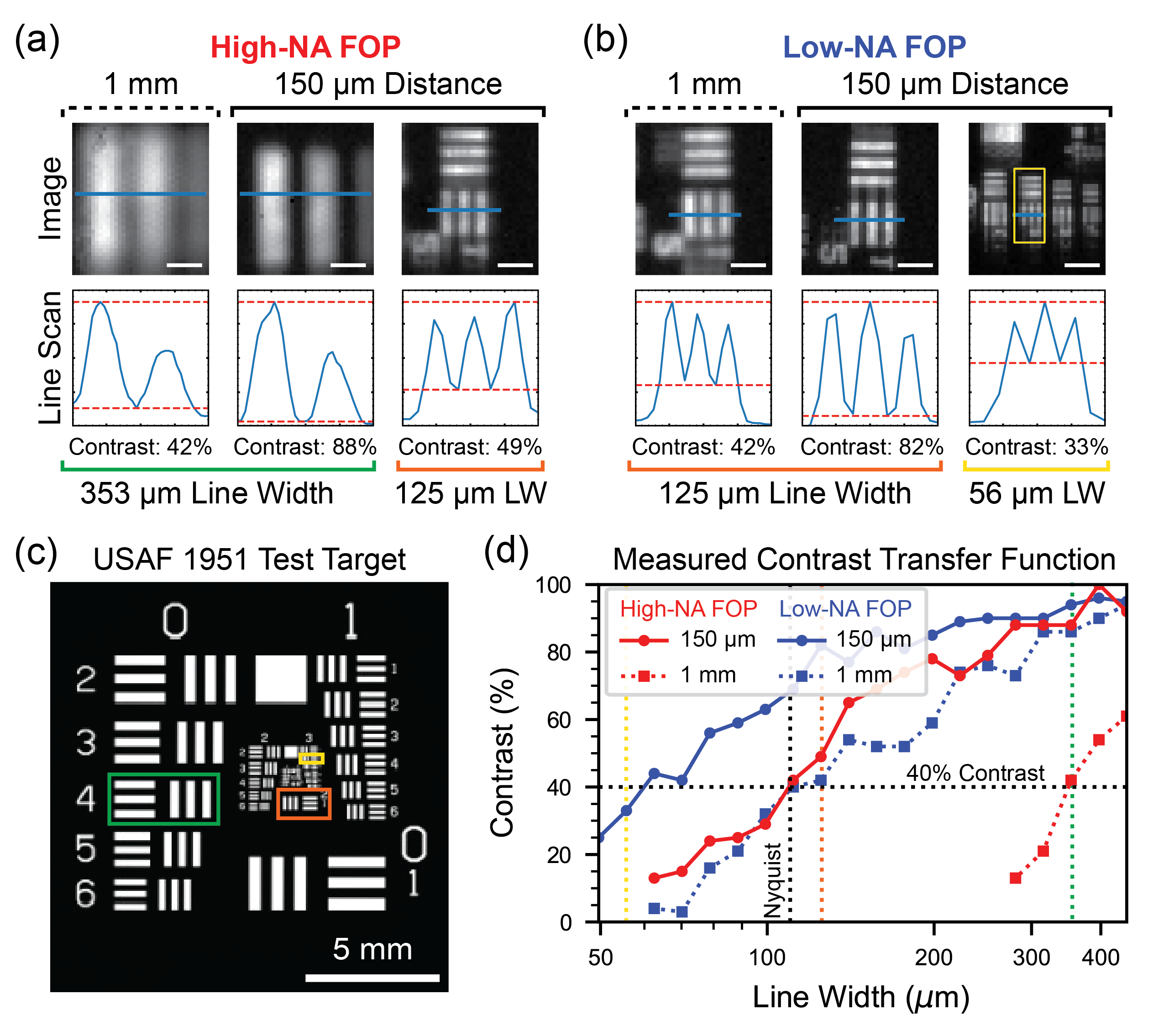}
\caption{\textbf{Resolution measurements. } Representative images and line scans of elements taken with (\textbf{a}) the high-NA FOP and (\textbf{b}) the low-NA FOP. Images are captured with a separation distance of both 150 $\mu$m and 1 mm between the test target and the sensor. The scale bar is 500 $\mu$m. (\textbf{c}) Reference image of 1951 USAF Test Target. Elements displayed in (a) and (b) are highlighted. (\textbf{d}) The measured contrast transfer for both FOPs at both separation distances.}
\label{resolution}
\end{figure}

Next, we measured the resolution of both designs by imaging a 1951 USAF target (R1L1S7N, Thorlabs). Fig. \ref{imaging setup}(b) shows a diagram of the imaging setup. The target is coated with IRDye 680LT at 10 $\mu$m concentration in DMSO and is sealed with a glass coverslip. The sample is imaged with the coverslip in contact with the sensor. A 635 nm fiber-coupled, collimated laser trans-illuminates the target, patterning the features onto the uniform layer of dye. To illustrate the effect of the working distance on the resolution, the experiment is repeated with coverslips that are 150-$\mu$m and 1-mm thick. For imaging, we use the 500-$\mu$m-thick high- and low-NA FOPs.
 
For both FOP designs and each coverslip thickness, we systematically image elements on the target with decreasing line widths (refer to Fig. \ref{resolution}(c)). At each position, 20 images are captured with an exposure time of 16 ms and are averaged to improve SNR. Fig. \ref{resolution}(a) and 8(b) show representative images for the high- and low-NA FOPs, respectively, with line scans through the center of each element showing the resulting contrast. As the line widths approach the resolution limit of the sensor, they blur together, reducing the contrast between the white and black bars. For quantification, we use Michelson contrast, defined as $\frac{I_{max}-I_{min}}{I_{max}+I_{min}}$, where $I_{max}$ is the brightest pixel value and $I_{min}$ is the darkest pixel value in the element. The calculated contrast transfer functions (CTFs) for each measurement are plotted in Fig. \ref{resolution}(d).

The resolution limit of the sensor is determined by both the pixel pitch and the optical resolution of the FOP. Given the 55-$\mu$m pixel pitch, the Nyquist-limited resolution of the sensor is 110 $\mu$m. For our measurements, we choose a contrast of 40\% as a basis for comparison. With the 150 $\mu$m coverslip, the resolution of both the high- and low-NA sensors is clearly Nyquist-limited. However, as the distance increases to 1 mm, the resolution degrades.  The low- and high-NA sensors resolve 109-$\mu$m and 349-$\mu$m line widths at 40\% contrast, respectively, a roughly 3.2× difference in resolution between the two designs. Accounting for the measured FWHMs in Fig. \ref{fop char}(b) and assuming the separating media is glass, the expected difference in resolution between the two designs is 4.7×. This difference is not fully realized in our measurements because with the low-NA FOP the resolution is still partially limited by the pixel pitch.

\subsection{Multicolor imaging}

\begin{figure}[tbp]
\centering\includegraphics[]{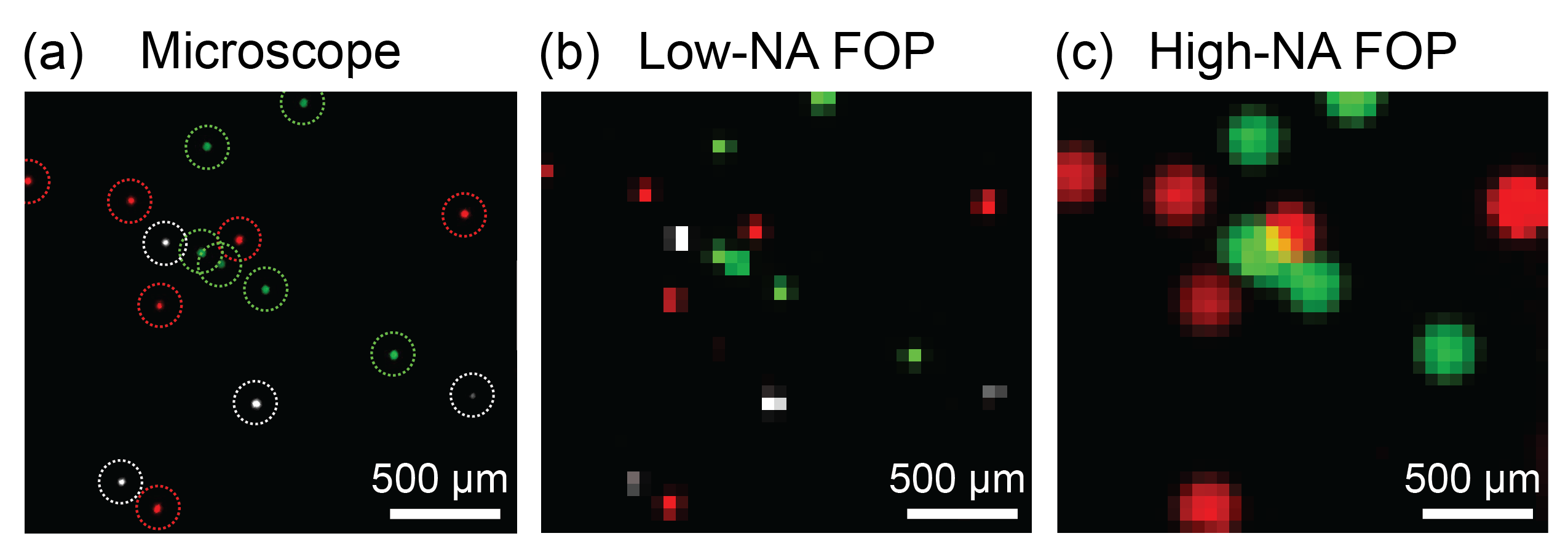}
\caption{\textbf{Multicolor imaging of fluorescent beads. } Image taken with (\textbf{a}) a microscope and a sensor with (\textbf{b}) the low-NA FOP and (\textbf{c}) the high-NA FOP.}
\label{multicolor}
\end{figure}

We demonstrate multicolor fluorescent imaging with a sample containing a mixture of 15-$\mu$m-diameter green ($\lambda_{EX}$=505 nm, $\lambda_{EM}$=515 nm, F8844, Thermo Fisher Scientific Inc.) and red (see Section 4.1) beads and 20-$\mu$m-diameter near-infrared (NIR) beads ($\lambda_{EX}$=780 nm, $\lambda_{EM}$=820 nm, 942774, Sigma-Aldrich) at roughly equivalent concentrations. The sample preparation is the same as in Section 4.1. We image the same sample using both the 500-$\mu$m-thick low- and high-NA FOPs. With the low-NA FOP, we use 488 nm, 660 nm, and 780 nm fiber-coupled lasers (QFLD-488-20S, QFLD-660-10S, and QFLD-780-50S, QPhotonics) for excitation operated at 3.8, 3.8, and 19.8 mW/cm\textsuperscript{2}, respectively. Since the high-NA FOP requires laser wavelengths further from the filter band-edge to achieve similar blocking, we use 455 nm and 635 nm fiber-coupled lasers (QFLD-450-10S and QFLD-635-40S, QPhotonics) operated 3.8 mW/cm\textsuperscript{2}. The NIR channel is incompatible with the high-NA FOP, because the filter notch cannot accommodate a laser wavelength shorter than 780 nm. 

The lasers are turned on individually and images of each color channel are captured with an exposure time of 32 ms. For each channel, 20 images are averaged, threshold-ed, false colored, and then overlaid to produce the final, multicolor image. Fig. \ref{multicolor}(a-c) shows the images of the bead sample captured with a fluorescence microscope (Leica DM IRB) and sensors with the low- and high-NA FOP, respectively. While the low-NA FOP achieves pixel-limited resolution, the resolution is lower with the high-NA FOP. 

\bgroup
\def\arraystretch{1}
 \begin{table}[htbp]
	\caption{\textbf{Comparison of optical frontend designs}}
	\label{tab:result summary}
 \centering
 \small
\renewcommand{\tabularxcolumn}[1]{>{\centering\arraybackslash}m{#1}}
\begin{tabularx}{1\textwidth}{ 
    >{\centering\arraybackslash}m{1.4cm} | 
    >{\centering\arraybackslash}c | 
    >{\centering\arraybackslash}c |
    >{\centering\arraybackslash}c | 
    >{\centering\arraybackslash}c |
    >{\centering\arraybackslash}c | 
    >{\centering\arraybackslash}c |
    X } 
\hline
\multirow{2}{*}{\textbf{Design}} &
\multirow{2}{*}{\textbf{$\boldsymbol{\theta_{FWHM}}$}} & 
\multirow{2}{*}{\textbf{\makecell{$\boldsymbol{0^\circ}$ T at \\ 660 nm}}} & 
\multirow{2}{*}{\textbf{Rejection}} & 
\multirow{2}{*}{\textbf{\makecell{Relative \\ CE}}} & 
\multicolumn{2}{c|}{\textbf{\makecell{Resolution at \\ 40\% Contrast}}} &
\multirow{2}{*}{\textbf{Colors}}\\
\cline{6-7}
& & & & & \textbf{\makecell{150 $\mu$m}} & \textbf{\makecell{1 mm}} & \\
\hline
\textbf{500 $\mu$m Low-NA} & $8.3^\circ$ &
27\% & > OD 7 & 1× & <110 $\mu$m & 109 $\mu$m & 3\\
\hline
\textbf{500 $\mu$m High-NA} & $45.7^\circ$ &
49\% & > OD 6 & 59× & <110 $\mu$m & 349 $\mu$m & 2\\
\end{tabularx}

\end{table}
\egroup

\section{Discussion}
Here we thoroughly analyze and improve on a technique using a FOP to compensate for the angle-dependent response of interference filters to enable multicolor, lensless fluorescence imaging. In particular, we show how the NA of the FOP can be increased to dramatically improve collection efficiency at a cost to resolution. Through our simulations and measurements, we develop a design methodology to accommodate larger NAs through two approaches. First, off-axis absorption of the FOP can be increased by increasing the thickness and reducing the fill factor. In general, increasing the thickness is preferred over adjusting the fill factor, which reduces the normal incidence transmittance. Second, the excitation wavelength can be moved further from the band-edge such that the interference filter is tolerant to a wider range of angles. However, this approach results in reduced fluorescence flux as the fluorophore is excited further from its absorption peak. While we implement low- and high-NA designs at the extremes of this design space, this design methodology can enable optical frontends with arbitrary intermediate NAs to trade-off collection efficiency and resolution according to sensor specifications. In addition, the NA of the FOP can be further increased by reducing the angle-sensitivity of the interference filter. This can be achieved by using high-index materials \cite{khurgin_expanding_2022} or by leveraging light-matter coupling \cite{mischok_breaking_2024}.

Our measurement results comparing the low- and high-NA designs are summarized in Table 1. The new high-NA design achieves 59× better collection efficiency then the low-NA design with 37× more fluorescence flux when accounting for off-peak absorption of the fluorophore. This significant improvement in collection efficiency is particularly useful for sensors with extremely low-photon flux, such as those that operate on stringent power budgets or need to detect ultra-low concentrations. The wider range of accepted angles could also enable computational imaging techniques to improve resolution and extract depth information \cite{boominathan_recent_2022}. Even without these techniques, the high-NA design offers pixel-limited resolution for working distances less than 150 $\mu$m, which are possible with trans-illuminated sensors. However, with the epi-illumination optics necessary for \textit{in vivo} imaging, the working distance increases and the resolution rapidly degrades with this design. While resolution may not be a concern for sensing applications, it is critical for imaging. 

Therefore, for imagers with longer working distances, the low-NA design may be preferred. We show that at a 1 mm working distance, the low-NA FOP achieves a resolution of at least 110 $\mu$m, which is at least 3.2× better than high-NA FOP. Given that the resolution is limited by the pixel pitch of our sensor, we calculate that this difference could be as large as 4.7× with a higher resolution image sensor. To be sure, computational imaging using a phase mask or diffuser has enabled less than 10 $\mu$m resolution at working distances of at least 3 mm \cite{kuo_-chip_2020,adams_vivo_2022} and with higher collection efficiencies. However, these rely on micro-fabricated optical components that increase manufacturing complexity and cost and require millimeter-scale working distances to operate, significantly increasing device thickness. They also use hand-optimized reconstruction algorithms that assume apriori constraints on the image sparsity such that they are prone to artifacts. Therefore, using the FOP as a resolving optic could be the preferred approach for reliable and compact imagers, especially for \textit{in vivo} imaging. A final advantage of the low-NA FOP, is that it operates with laser wavelengths less than 20 nm from the band-edge. This configuration enables tightly spaced passbands for highly multiplexed fluorescence imaging. 

In implementing these designs, we show that when the interference filter is coated on a single side of the FOP, as was done in our prior work, scattering reduces the rejection to as little as OD 3.7 at certain AOIs depending on the relative orientation of the filter and the FOP. To address this issue, we demonstrated a new design in which the interference filter is coated on both sides of the FOP, achieving at least OD 6 across all AOIs. In most cases, this dual-sided design is preferred because it is tolerant to scattering in the sample. However, the single-sided configurations are adequate in situations where the AOI of the excitation light is well-controlled and can offer savings in manufacturing cost and complexity. For example, the filter-first configuration is compatible with sensors using trans-illumination where the excitation light is collimated and at normal incidence. On the other hand, the filter-last configuration is best suited to scenarios in which the AOI of the excitation light is primarily oblique, such as in devices using epi-illumination optics with edge-coupled sources. Another advantage of these single-sided designs is that they can be thinned on their backside after coating to achieve a thinner device, while the dual-sided design is constrained by the minimum 500 $\mu$m substrate thickness in our coating process. To this end, we show that the low-NA FOP can be thinned to 250 $\mu$m, while still providing at least OD 6 rejection under controlled illumination.

Table \ref{comparison table} compares the performance of our low- and high-NA optical frontend designs with recently published filters for lensless fluorescence sensors \cite{yildirim_implementation_2017,papageorgiou_angle-insensitive_2018,sasagawa_highly_2018,moazeni_mechanically_2021,kulmala_lensless_2022,taal_toward_2022,pollmann_subdural_2024,marin-lizarraga_simultaneous_2025,aghlmand_65-nm_2023,hong_nano-plasmonics_2018,shin_miniaturized_2023,choi_fully_2020}. Although dual-color imaging has been demonstrated with hybrid filters (interference and absorption), our design is the only to achieve three-color imaging and the only capable of imaging across both the visible and NIR spectrum. This unlocks the full range of commonly used fluorophores. While visible fluorophores such as green fluorescent protein (GFP) \cite{tsien_green_1998} are the standard in laboratory science, NIR fluorophores are preferred for \textit{in vivo} imaging due to reduced scattering, absorption, and autofluorescence at longer wavelengths \cite{frangioni_vivo_2003}. Moreover, our designs also offer at least OD 6 rejection, which is competitive with the best hybrid filter designs. Also, the low-NA design has a roll-off of only 17 nm (the best to our knowledge), demonstrating optimal compatibility with the 10–30 nm Stokes shifts of common organic fluorophores and ideal for highly multiplexed imaging. The 20–50\% pass-band transmittance with our designs falls within the range of other reported filters. While thinner designs are possible with other techniques, the thickness of the FOP does not have a significant effect on resolution, as TIR within the fibers prevents the emission light from diverging. 

Overall, we demonstrate an exceptionally versatile optical frontend for lensless, multicolor fluorescence imaging that is adaptable to different fluorophores, illumination configurations, and lensless imaging techniques. The design is fully fabricated using widely available commercial components and services and can leverage the vast existing catalog of commercial interference filters designed for conventional fluorescence microscopes. On the other hand, absorption filters require unique materials for each wavelength, significantly limiting design versatility and multiplexed imaging. Therefore, we have observed that the majority of published research in lensless fluorescence sensing has devoted significant effort to developing bespoke filters as a prerequisite for research into other aspects of sensor design. With this work, we aim to offer a versatile plug-and-play optical frontend to enable new research into high-performance CMOS sensor architectures, lensless imaging techniques, and novel applications for miniaturized fluorescence sensors.  

\vspace{8 pt}

\textbf{Funding.}
{\small{National Cancer Institute (1R01CA290027-01); National Institute of Dental and Craniofacial Research (1DP2DE030713-01); Advanced Research Projects Agency for Health (ARPA-H) (D24AC00342-00); National Cancer Institute (5R01CA278672-02).}}

\vspace{8 pt}

\textbf{Acknowledgments.}
{\small{The authors thank Chroma Technologies Corp. for assistance with custom filter depositions and the sponsors of BSAC (Berkeley Sensors and Actuators Center) and TSMC for chip fabrication. We thank the University of California, Berkeley Marvell Nanofabrication Laboratory for support with dicing the FOPs. We also thank Prof. Peter Hosemann and Kook Noh Yoon for help with lapping and polishing. Research reported in this publication was supported by the Advanced Research Projects Agency for Health (ARPA-H) under Award Number D24AC00342-00. This ARPA-H award provided 70\% of the total costs with the other 30\% from other federal funding sources. The content is solely the responsibility of the authors and does not necessarily represent the official views of the Advanced Research Projects Agency for Health. The contents are those of the author. They may not reflect the policies of the Department of Health and Human Services or the U.S. government. The content is solely the responsibility of the authors and does not necessarily represent the official views of the Advanced Research Projects Agency for Health. }}

\vspace{8 pt}

\textbf{Disclosures.} {\small{Micah Roschelle and Mekhail Anwar applied for a patent entitled “Apparatus, Systems, and Methods for Fluorescence Imaging on Planar Sensors” related to this paper. Mekhail Anwar is a shareholder in Insyte Bio.}}

\vspace{8 pt}

\textbf{Data availability.}{\small{ Data underlying the results presented in this paper are not publicly available at this time but may be obtained from the authors upon reasonable request.}}



\begin{thebibliography}{10}
\newcommand{\enquote}[1]{``#1''}

\bibitem{greenbaum_imaging_2012}
A.~Greenbaum, W.~Luo, T.~W. Su, \emph{et~al.}, \enquote{Imaging without lenses: {Achievements} and remaining challenges of wide-field on-chip microscopy,} {\protect\JournalTitle{Nature Methods}} \textbf{9}, 889--895 (2012).

\bibitem{boominathan_recent_2022}
V.~Boominathan, J.~T. Robinson, L.~Waller, and A.~Veeraraghavan, \enquote{Recent advances in lensless imaging,} {\protect\JournalTitle{Optica}} \textbf{9}, 1 (2022). Publisher: The Optical Society.

\bibitem{wu_bio-flatscope_2021}
J.~Wu, D.~Yan, V.~Boominathan, \emph{et~al.}, \enquote{Bio-{FlatScope}: a flat, lensless microscope for fluorescence imaging,} Tech. rep. (2021).

\bibitem{pollmann_subdural_2024}
E.~H. Pollmann, H.~Yin, I.~Uguz, \emph{et~al.}, \enquote{A subdural {CMOS} optical device for bidirectional neural interfacing,} {\protect\JournalTitle{Nature Electronics}} pp. 1--13 (2024). Publisher: Nature Publishing Group.

\bibitem{rabbani_towards_2024}
R.~Rabbani, H.~Najafiaghdam, M.~Roschelle, \emph{et~al.}, \enquote{Towards {A} {Wireless} {Image} {Sensor} for {Real}-{Time} {Fluorescence} {Microscopy} in {Cancer} {Therapy},} {\protect\JournalTitle{IEEE Transactions on Biomedical Circuits and Systems}} pp. 1--15 (2024). Conference Name: IEEE Transactions on Biomedical Circuits and Systems.

\bibitem{roschelle_wireless_2024}
M.~Roschelle, R.~Rabbani, S.~Gweon, \emph{et~al.}, \enquote{A {Wireless}, {Multicolor} {Fluorescence} {Image} {Sensor} {Implant} for {Real}-{Time} {Monitoring} in {Cancer} {Therapy},} {\protect\JournalTitle{IEEE Journal of Solid-State Circuits}} pp. 1--19 (2024). Conference Name: IEEE Journal of Solid-State Circuits.

\bibitem{zhu_ingestible_2023}
C.~Zhu, Y.~Wen, T.~Liu, \emph{et~al.}, \enquote{An {Ingestible} {Pill} {With} {CMOS} {Fluorescence} {Sensor} {Array}, {Bi}-{Directional} {Wireless} {Interface} and {Packaged} {Optics} for in-{Vivo} {Bio}-{Molecular} {Sensing},} {\protect\JournalTitle{IEEE Transactions on Biomedical Circuits and Systems}} \textbf{17}, 257--272 (2023). Publisher: Institute of Electrical and Electronics Engineers Inc.

\bibitem{papageorgiou_chip-scale_2020}
E.~P. Papageorgiou, B.~E. Boser, and M.~Anwar, \enquote{Chip-{Scale} {Angle}-{Selective} {Imager} for in {Vivo} {Microscopic} {Cancer} {Detection},} {\protect\JournalTitle{IEEE Transactions on Biomedical Circuits and Systems}} \textbf{14}, 91--103 (2020). Publisher: Institute of Electrical and Electronics Engineers Inc.

\bibitem{roschelle_multicolor_2024}
M.~Roschelle, R.~Rabbani, E.~Papageorgiou, \emph{et~al.}, \enquote{Multicolor fluorescence microscopy for surgical guidance using a chip-scale imager with a low-{NA} fiber optic plate and a multi-bandpass interference filter,} {\protect\JournalTitle{Biomedical Optics Express}} \textbf{15}, 1761 (2024).

\bibitem{xiahou_-chip_2025}
Y.~Xiahou, B.~Wang, H.~Li, \emph{et~al.}, \enquote{On-{Chip} {Array} {Fluorescent} {Sensor} for {High}-{Sensitivity} {Multi}-{Gas} {Detection},} {\protect\JournalTitle{ACS Sensors}} \textbf{10}, 3647--3657 (2025). Publisher: American Chemical Society.

\bibitem{aghlmand_65-nm_2023}
F.~Aghlmand, C.~Y. Hu, S.~Sharma, \emph{et~al.}, \enquote{A 65-nm {CMOS} {Fluorescence} {Sensor} for {Dynamic} {Monitoring} of {Living} {Cells},} {\protect\JournalTitle{IEEE Journal of Solid-State Circuits}} \textbf{58}, 3003--3019 (2023).

\bibitem{dandin_optical_2007}
M.~Dandin, P.~Abshire, and E.~Smela, \enquote{Optical filtering technologies for integrated fluorescence sensors,} {\protect\JournalTitle{Lab on a Chip}} \textbf{7}, 955--977 (2007). Publisher: Royal Society of Chemistry.

\bibitem{yildirim_implementation_2017}
E.~Yildirim, C.~Arpali, and S.~A. Arpali, \enquote{Implementation and characterization of an absorption filter for on-chip fluorescent imaging,} {\protect\JournalTitle{Sensors and Actuators, B: Chemical}} \textbf{242}, 318--323 (2017). Publisher: Elsevier B.V.

\bibitem{papageorgiou_angle-insensitive_2018}
E.~P. Papageorgiou, H.~Zhang, B.~E. Boser, \emph{et~al.}, \enquote{Angle-insensitive amorphous silicon optical filter for fluorescence contact imaging,} {\protect\JournalTitle{Optics Letters}} \textbf{43}, 354 (2018). Publisher: The Optical Society.

\bibitem{moazeni_mechanically_2021}
S.~Moazeni, E.~Pollmann, V.~Boominathan, \emph{et~al.}, \enquote{A {Mechanically} {Flexible}, {Implantable} {Neural} {Interface} for {Computational} {Imaging} and {Optogenetic} {Stimulation} over 5.4$\times$5.4mm\textsuperscript{2} {FoV},} {\protect\JournalTitle{IEEE Transactions on Biomedical Circuits and Systems}} \textbf{15}, 1295--1305 (2021). Publisher: Institute of Electrical and Electronics Engineers Inc.

\bibitem{sasagawa_highly_2018}
K.~Sasagawa, A.~Kimura, M.~Haruta, \emph{et~al.}, \enquote{Highly sensitive lens-free fluorescence imaging device enabled by a complementary combination of interference and absorption filters,} {\protect\JournalTitle{Biomedical Optics Express}} \textbf{9}, 4329 (2018). Publisher: The Optical Society.

\bibitem{marin-lizarraga_simultaneous_2025}
V.~Marin-Lizarraga, R.~Rodriguez-Garcia, J.~L. Garcia-Cordero, \emph{et~al.}, \enquote{Simultaneous bright-field and fluorescence lensless imaging with high excitation light extinction for microfluidics applications,} {\protect\JournalTitle{Optics and Lasers in Engineering}} \textbf{185}, 108724 (2025).

\bibitem{kulmala_lensless_2022}
N.~Kulmala, K.~Sasagawa, T.~Treepetchkul, \emph{et~al.}, \enquote{Lensless dual-color fluorescence imaging device using hybrid filter,} {\protect\JournalTitle{Japanese Journal of Applied Physics}} \textbf{61} (2022). Publisher: IOP Publishing Ltd.

\bibitem{taal_toward_2022}
A.~J. Taal, C.~Lee, J.~Choi, \emph{et~al.}, \enquote{Toward implantable devices for angle-sensitive, lens-less, multifluorescent, single-photon lifetime imaging in the brain using Fabry Perot and absorptive color filters,} {\protect\JournalTitle{Light: Science and Applications}} \textbf{11} (2022). Publisher: Springer Nature.

\bibitem{hong_nano-plasmonics_2018}
L.~Hong, H.~Li, H.~Yang, and K.~Sengupta, \enquote{Nano-plasmonics and electronics co-integration in {CMOS} enabling a pill-sized multiplexed fluorescence microarray system,} {\protect\JournalTitle{Biomedical Optics Express}} \textbf{9}, 5735--5758 (2018). Publisher: Optica Publishing Group.

\bibitem{shin_miniaturized_2023}
H.~Shin, G.-W. Yoon, W.~Choi, \emph{et~al.}, \enquote{Miniaturized multicolor fluorescence imaging system integrated with a {PDMS} light-guide plate for biomedical investigation,} {\protect\JournalTitle{npj Flexible Electronics}} \textbf{7}, 7 (2023).

\bibitem{choi_fully_2020}
J.~Choi, A.~J. Taal, W.~L. Meng, \emph{et~al.}, \enquote{Fully {Integrated} {Time}-{Gated} {3D} {Fluorescence} {Imager} for {Deep} {Neural} {Imaging},} {\protect\JournalTitle{IEEE Transactions on Biomedical Circuits and Systems}} \textbf{14}, 636--645 (2020). Publisher: Institute of Electrical and Electronics Engineers Inc.

\bibitem{najafiaghdam_optics-free_2022}
H.~Najafiaghdam, C.~C. Pedroso, B.~E. Cohen, and M.~Anwar, \enquote{Optics-{Free} {Chip}-{Scale} {Intraoperative} {Imaging} {Using} {NIR}-{Excited} {Upconverting} {Nanoparticles},} {\protect\JournalTitle{IEEE Transactions on Biomedical Circuits and Systems}} \textbf{16}, 312--323 (2022). Publisher: Institute of Electrical and Electronics Engineers Inc.

\bibitem{berezin_fluorescence_2010}
M.~Y. Berezin and S.~Achilefu, \enquote{Fluorescence {Lifetime} {Measurements} and {Biological} {Imaging},} {\protect\JournalTitle{Chemical Reviews}} \textbf{110}, 2641--2684 (2010). Publisher: American Chemical Society.

\bibitem{sasagawa_front-light_2023}
K.~Sasagawa, Y.~Ito, D.~Schaeffer, \emph{et~al.}, \enquote{Front-light {Structure} for a {Lensless} {Fluorescence} {Imaging} {Device} with a {Hybrid} {Emission} {Filter},} in \emph{2023 {IEEE} {Biomedical} {Circuits} and {Systems} {Conference} ({BioCAS}),}  (2023), pp. 1--5. ISSN: 2766-4465.

\bibitem{adams_vivo_2022}
J.~K. Adams, D.~Yan, J.~Wu, \emph{et~al.}, \enquote{In vivo lensless microscopy via a phase mask generating diffraction patterns with high-contrast contours,} {\protect\JournalTitle{Nature Biomedical Engineering}}  (2022).

\bibitem{antipa_diffusercam_2018}
N.~Antipa, G.~Kuo, R.~Heckel, \emph{et~al.}, \enquote{{DiffuserCam}: lensless single-exposure {3D} imaging,} {\protect\JournalTitle{Optica}} \textbf{5}, 1 (2018). ArXiv: 1710.02134 Publisher: The Optical Society.

\bibitem{kuo_-chip_2020}
G.~Kuo, F.~L. Liu, I.~Grossrubatscher, \emph{et~al.}, \enquote{On-chip fluorescence microscopy with a random microlens diffuser,} {\protect\JournalTitle{Optics Express}} \textbf{28}, 8384--8399 (2020). Publisher: Optica Publishing Group.

\bibitem{balsam_lensless_2011}
J.~Balsam, M.~Ossandon, Y.~Kostov, \emph{et~al.}, \enquote{Lensless {CCD}-based fluorometer using a micromachined optical {S{\"o}ller} collimator,} {\protect\JournalTitle{Lab on a Chip}} \textbf{11}, 941--949 (2011). Publisher: Royal Society of Chemistry.

\bibitem{xia_wave_2020}
M.~M. Xia, B.~Walter, E.~Michielssen, \emph{et~al.}, \enquote{A wave optics based fiber scattering model,} {\protect\JournalTitle{ACM Transactions on Graphics}} \textbf{39}, 1--16 (2020).

\bibitem{rabbani_173_2024}
R.~Rabbani, M.~Roschelle, S.~Gweon, \emph{et~al.}, \enquote{17.3 {A} {Fully} {Wireless}, {Miniaturized}, {Multicolor} {Fluorescence} {Image} {Sensor} {Implant} for {Real}-{Time} {Monitoring} in {Cancer} {Therapy},} in \emph{2024 {IEEE} {International} {Solid}-{State} {Circuits} {Conference} ({ISSCC}),}  vol.~67 (2024), pp. 318--320. ISSN: 2376-8606.

\bibitem{khurgin_expanding_2022}
J.~B. Khurgin, \enquote{Expanding the {Photonic} {Palette}: {Exploring} {High} {Index} {Materials},} {\protect\JournalTitle{ACS Photonics}} \textbf{9}, 743--751 (2022). Publisher: American Chemical Society.

\bibitem{mischok_breaking_2024}
A.~Mischok, B.~Siegmund, F.~Le~Roux, \emph{et~al.}, \enquote{Breaking the angular dispersion limit in thin film optics by ultra-strong light-matter coupling,} {\protect\JournalTitle{Nature Communications}} \textbf{15}, 10529 (2024). Publisher: Nature Publishing Group.

\bibitem{tsien_green_1998}
R.~Y. Tsien, \enquote{{THE} {GREEN} {FLUORESCENT} {PROTEIN},} {\protect\JournalTitle{Annual Review of Biochemistry}} \textbf{67}, 509--544 (1998). Publisher: Annual Reviews.

\bibitem{frangioni_vivo_2003}
J.~V. Frangioni, \enquote{\textit{{In} vivo} near-infrared fluorescence imaging,} {\protect\JournalTitle{Current Opinion in Chemical Biology}} \textbf{7}, 626--634 (2003).

\end{thebibliography}

\newpage

\renewcommand{\thetable}{S\arabic{table}} 

\renewcommand{\thefigure}{S\arabic{figure}} 

\setcounter{section}{0}
\setcounter{figure}{0}
\setcounter{table}{0}
\setcounter{equation}{0}
\renewcommand{\theHequation}{S\theequation} 
\renewcommand{\theHsection}{S\thesection}

\title{A Versatile Optical Frontend for Multicolor Fluorescence Imaging with Miniaturized Lensless Sensors: Supplement}

\author{Lukas Harris\authormark{1,2,†}, Micah Roschelle\authormark{1,2,†,*}, Jack Bartley \authormark{2}, and Mekhail Anwar\authormark{1,2,*}}

\address{\authormark{1}Department of Electrical Engineering and Computer Sciences, University of California, Berkeley, California 94720, USA}
\address{\authormark{2}Department of Radiation Oncology, University of California, San Francisco, California 94158, USA}
\address{\authormark{†}These authors contributed equally to this work.}
\email{\authormark{*}micah.roschelle@berkeley.edu} \email{\authormark{*}mekhail.anwar@ucsf.edu}

\section{PSF of a lens-less imager}
\begin{figure}[htbp]
\centering\includegraphics[]{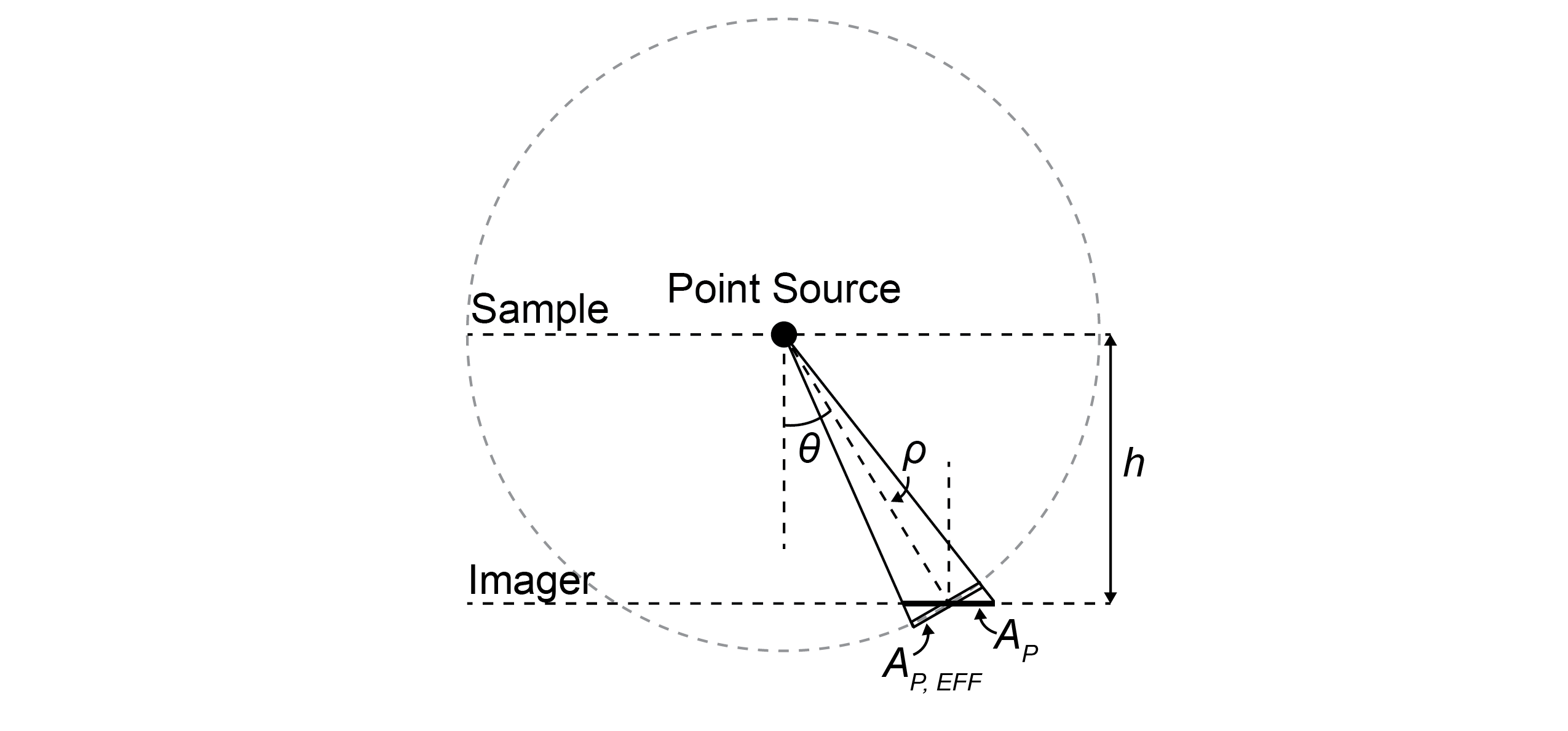}
\caption{2D cross-section of geometry used for deriving 2D PSF and collection efficiency of a general lens-less imager.}
\label{derivations}
\end{figure}

The point spread function, $PSF(\theta)$, of a general lens-less imaging system with angular transmittance function, $T(\theta)$, can be derived as follows. Fig. \ref{derivations} shows a 2D cross-section of the geometry used for the derivation. A point source is located a vertical distance, $h$, above the image sensor. $PSF(\theta)$ is found by determining the fraction of total light emitted by the point source that is captured by a small pixel on the image sensor located at radial distance, $\rho$, and angle, $\theta$, from the point source, where $\rho=\frac{h}{\cos\theta}$. 

The point source emits total optical power, $P_0=4\pi I_0$ as an isotropic flux, $\Phi$, through a spherical surface. As the light diverges with increasing $\rho$, $\Phi$ decays according to 
\begin{equation}
    \Phi=\frac{I_0}{\rho^2}=I_0 \cos^2 \theta
\end{equation}
The total integrated flux on the pixel at a given $\theta$ (which is equivalent to $PSF(\theta)$) depends on the area of the pixel, $A_P$, and the angular $T(\theta)$ of the sensor. To determine area of the sphere bounding subtended by the pixel, we calculate the effective area of the pixel, $A_{P, EFF}$, tangent to the sphere with radius $\rho$, as  $A_{P, EFF}=A_P \cdot \cos\theta$ . Therefore, if the pixel is very small relative to $h$, 
\begin{equation}
PSF(\theta) = T(\theta)\cdot \Phi\cdot A_{P, EFF}=T(\theta)\cdot I_0 \cos^2 \theta \cdot A_P \cdot \cos \theta \propto T(\theta) \cdot \cos^3 \theta 
\end{equation}
Here, we assume that small  $A_{P, EFF}$ is approximately a patch on the sphere and the pixel approximately subtends only a single $\theta$. 

Therefore, for a lensless imager without a collimator, the FWHM angular resolution, $\theta_{FWHM}=75^\circ$.  This can be improved by restricting $T(\theta)$ using a collimator.

\section{Collection efficiency of lens-less imager}
Collection efficiency, $\eta_C$ is defined as the fraction of the total light emitted from a point source that is captured by the imager. As in the preceding section, the point source emits with flux, $\Phi=\frac{I_0}{\rho^2}$. The amount of light captured by the sensor, $P_C$ is the integrated surface area subtended by the sensor weighted by $T(\theta)$:
\begin{equation}
      P_C=\frac{I_0}{\rho^2}\int T(\theta)\,dA =\frac{I_0}{\rho^2}\int_{0}^{2\pi} \int_{0}^{\dfrac{\pi}{2}} T(\theta) \rho^2 \sin\theta \,d\theta \,d\phi=2\pi I_0 \int_{0}^{\dfrac{\pi}{2}} T(\theta)\sin\theta \,d\theta
\end{equation}  

Here we assume that the image sensor is approximately an infinite plane with respect to the point source. We formulate this integral in a spherical coordinate system system where $\theta$ is the polar angle and $\phi$ is the azimuthal angle. 

To calculate the collection efficiency, we normalize by the total optical power of the point source $P_0=4\pi I_0$ such that,
\begin{equation}
    \eta_C=\frac{1}{2} \int_{0}^{\dfrac{\pi}{2}} T(\theta)\sin\theta \,d\theta
\end{equation}
For an ideal rectangular angle filter, $T(\theta)=\Pi(\frac{\theta}{\theta_C})$, the collection efficiency is
\begin{equation}
    \eta_C=\frac{1}{2} \int_0^{\theta_C} \sin\theta \,d\theta=\frac{1}{2}[1-\cos \theta_C]=\sin^2 \frac{\theta_C}{2}
\end{equation}

\section{Choosing fiber pitch}
\begin{figure}[htbp]
\centering\includegraphics[]{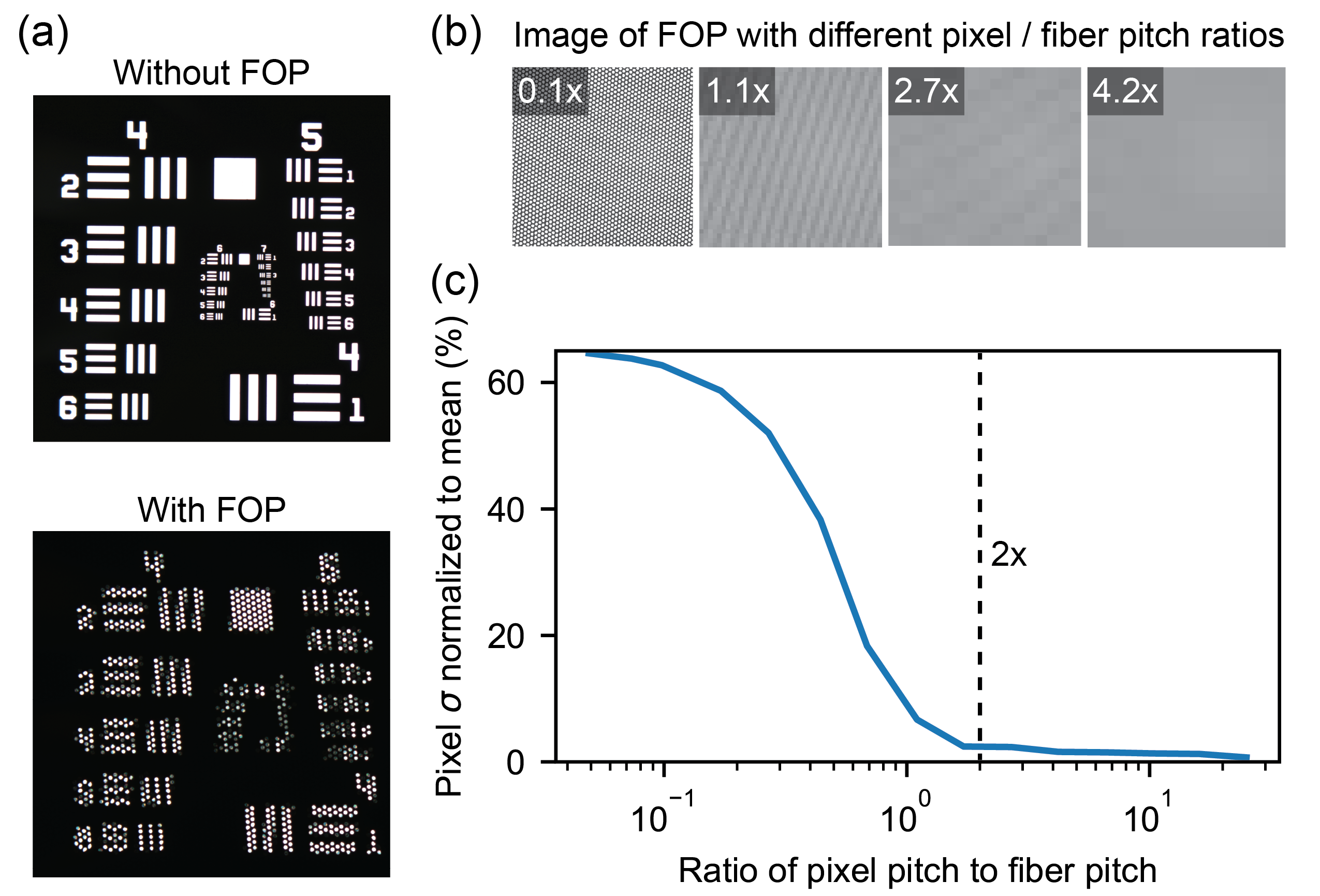}
\caption{\textbf{Effect of fiber pitch on resolution.} (\textbf{a}) Microscope images of a USAF target with and without the FOP over the target. The resolution is affected by the fiber pitch of the FOP.  (\textbf{b}) Image of the FOP sampled with different pixel sizes. A larger pixel-to-fiber pitch ratio leads to a more uniform image. (\textbf{c}) Quantification of pixel variation for different pixel to fiber pitch ratios.}
\label{fiber sizing}
\end{figure}

The choice of the fiber pitch of the FOP depends on the pixel pitch of the image sensor. As the transmitted light reflects within the fiber, the pitch of the fiber determines the best resolution possible with the FOP. This effect is illustrated in Fig. \ref{fiber sizing}(a) which shows a microscope image of a 1951 USAF test target taken with and without the FOP overlaid. Clearly, the FOP limits the resolution of the microscope well beyond the pixel-limited resolution. Therefore, for pixel-limited resolution, the fibers must be smaller than the pixels. 

Also, if the fibers are not exactly pitch-matched and aligned to the pixels, the responsivity of each pixel will be modulated differently depending on the percentage of the pixel area occupied by fiber cores. This effect is illustrated in Fig. \ref{fiber sizing}(b) where an image of the low-NA FOP is sampled with different pixel sizes. As the pixel pitch becomes larger than the fiber pitch, the image becomes more uniform. This effect is quantified for different pixel-pitch-to-fiber-pitch ratios in Fig. \ref{fiber sizing}(c). Here we show that the pixel standard deviation starts to reach a minimum with a ratio of 2×.

\begin{figure}[htbp]
\centering\includegraphics[]{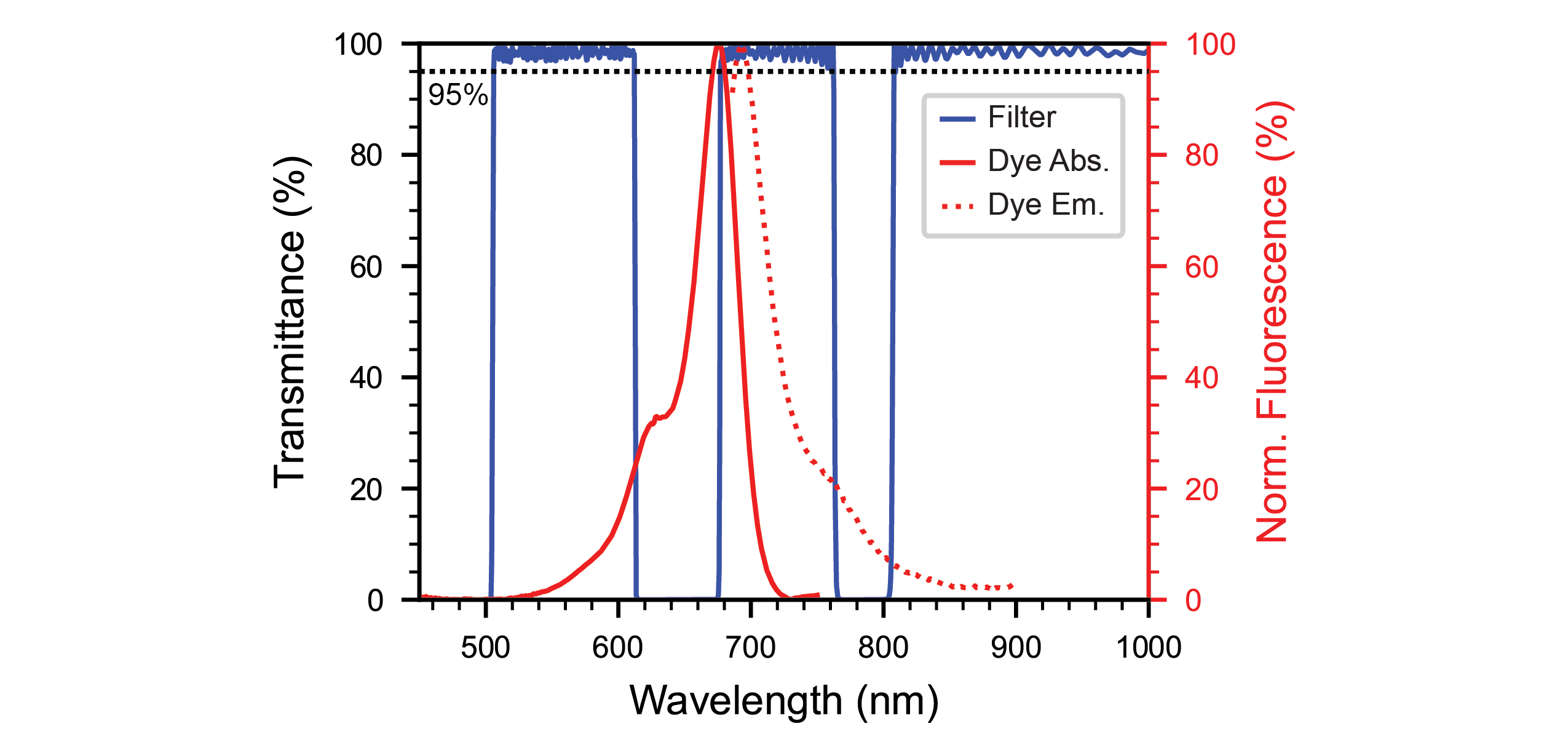}
\caption{\textbf{Filter transmittance spectra. }Transmittance spectra of the interference filter at normal incidence on a linear scale. The absorption and emission spectra of IRDye 680LT are overlaid. 
}
\label{lin filter spectra}
\end{figure}

\begin{figure}[htbp]
\centering\includegraphics[]{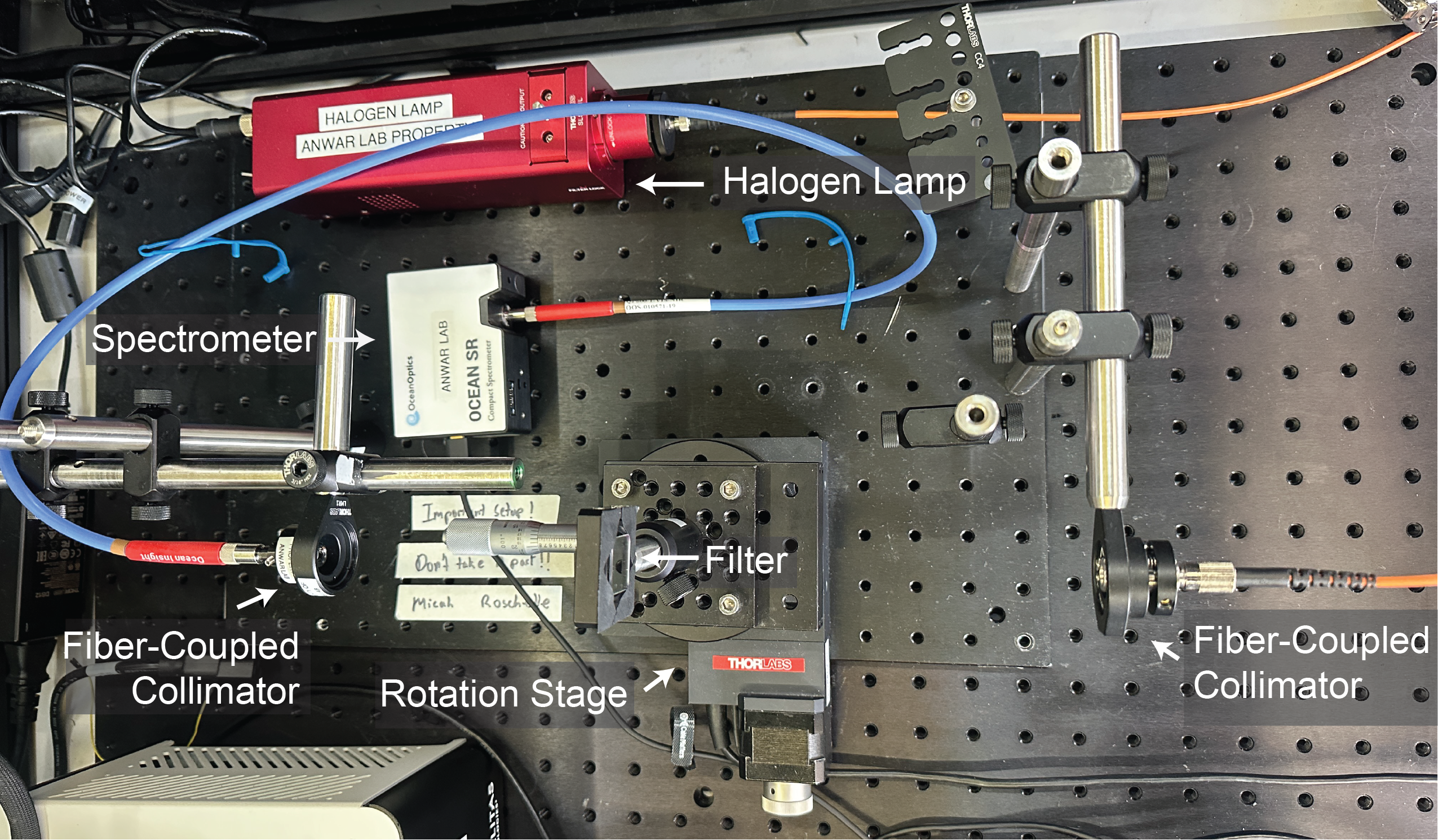}
\caption{\textbf{Setup for measuring spectral transmittance across AOIs.} 
The device under test is mounted on an automated rotation stage (HDR50, Thorlabs). A collimated, fiber-coupled tungsten-halogen lamp (SLS201L, Thorlabs) illuminates the device and the spectral transmittance is measured with a fiber-coupled spectrometer (SR-6, Ocean Optics). Transmittance spectra are captured at different angles as the stage rotates. 
}
\label{spectra setup}
\end{figure}

\begin{figure}[htbp]
\centering\includegraphics[]{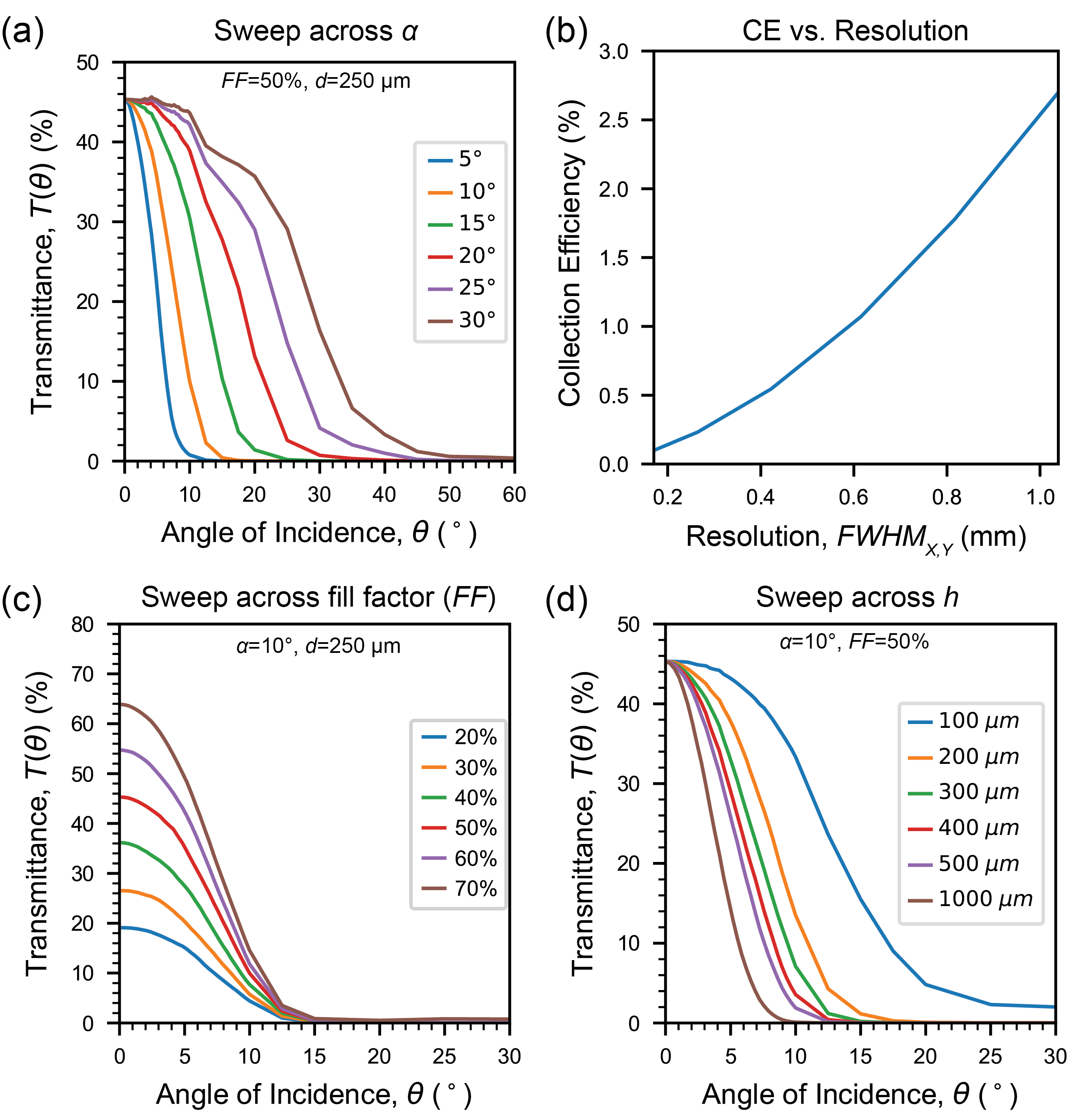}
\caption{\textbf{FOP Simulations.} (\textbf{a}) Simulated transmittance across AOIs for different fiber acceptance angles ($\alpha$) on a log scale. (\textbf{b}) Calculated collection efficiency vs. resolution ($FWHM_{X,Y}$) for the simulated FOPs with different NAs in (a). (\textbf{c}) Simulated transmittance across AOIs for different fill factors ($FF$) on a linear scale. (\textbf{d}) Simulated transmittance across AOIs for FOPs of different thicknesses ($h$) on a linear scale.  }
\label{lin sim results}
\end{figure}

\begin{figure}[htbp]
\centering\includegraphics[]{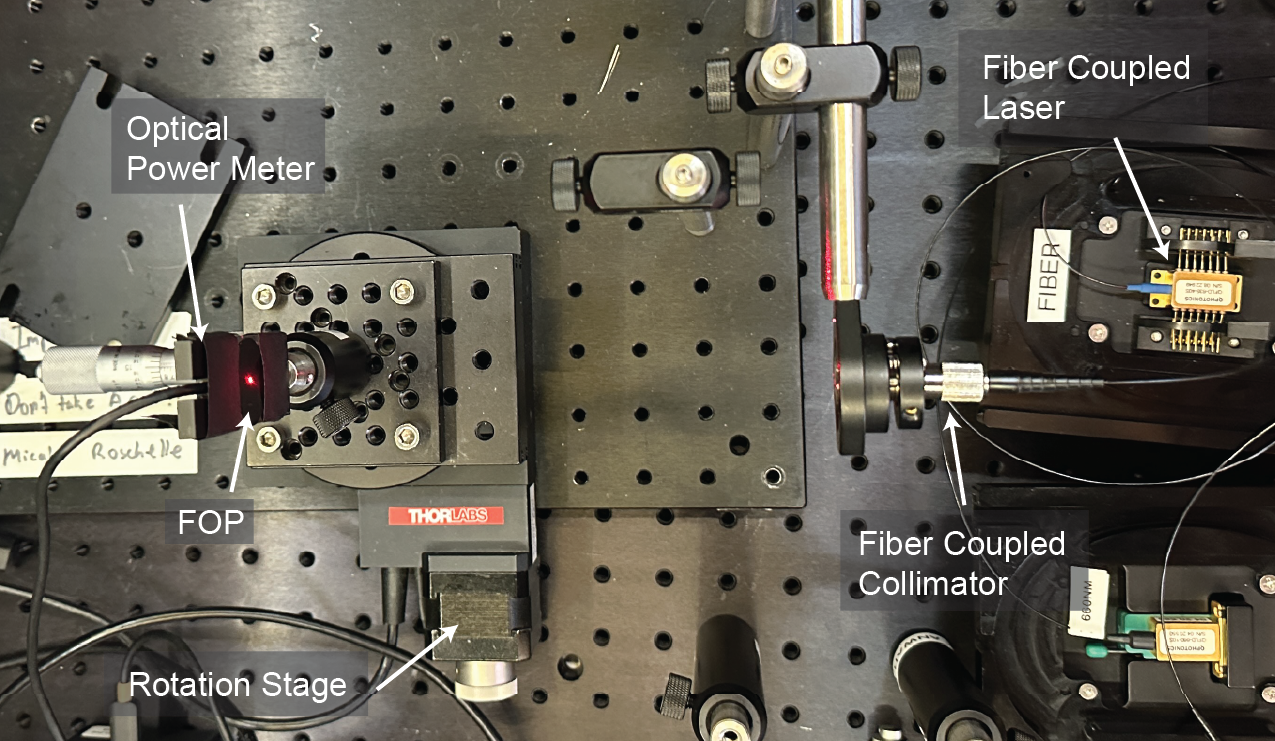}
\caption{\textbf{Angular Transmittance Measurement Setup.} The device under test is taped to the power meter photodiode (PM100D with S171C, Thorlabs) which is mounted on an motorized rotation station (HDR50, Thorlabs). The transmittance is measured with a collimated, fiber-coupled laser (QFLD-660-10S, QPhotonics). 
}
\label{trans meas setup}
\end{figure}

\begin{figure}[htbp]
\centering\includegraphics[]{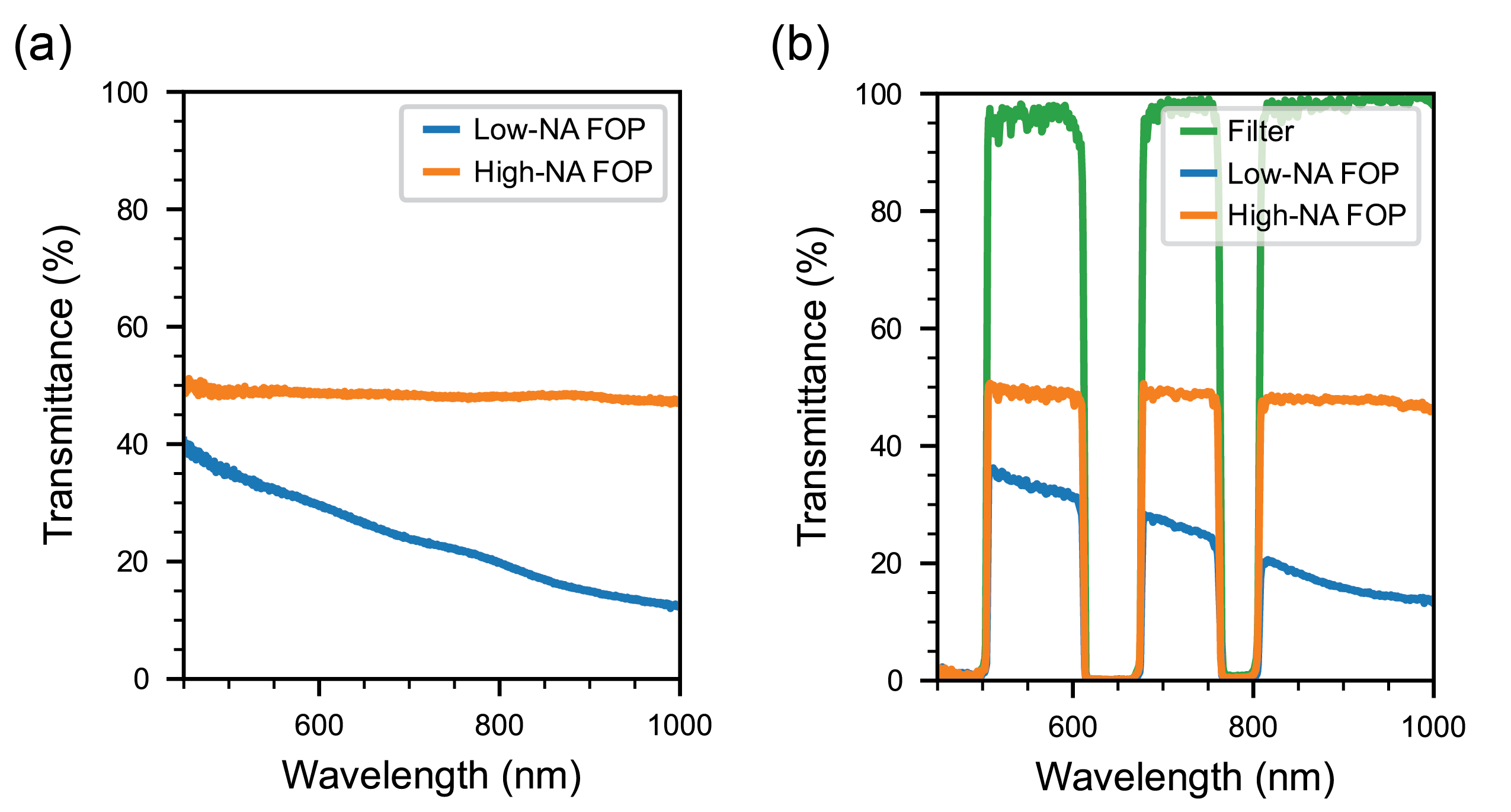}
\caption{\textbf{Optical frontend transmittance spectra at 0 degrees.} 
(\textbf{a}) Measured normal-incidence transmittance spectra of the low- and high-NA FOPs. (\textbf{b}) Measured normal-incidence transmittance spectra of optical frontend using the low- and high-NA FOPs in the filter first configuration. }
\label{trans spectra}
\end{figure}

\begin{figure}[htbp]
\centering\includegraphics[]{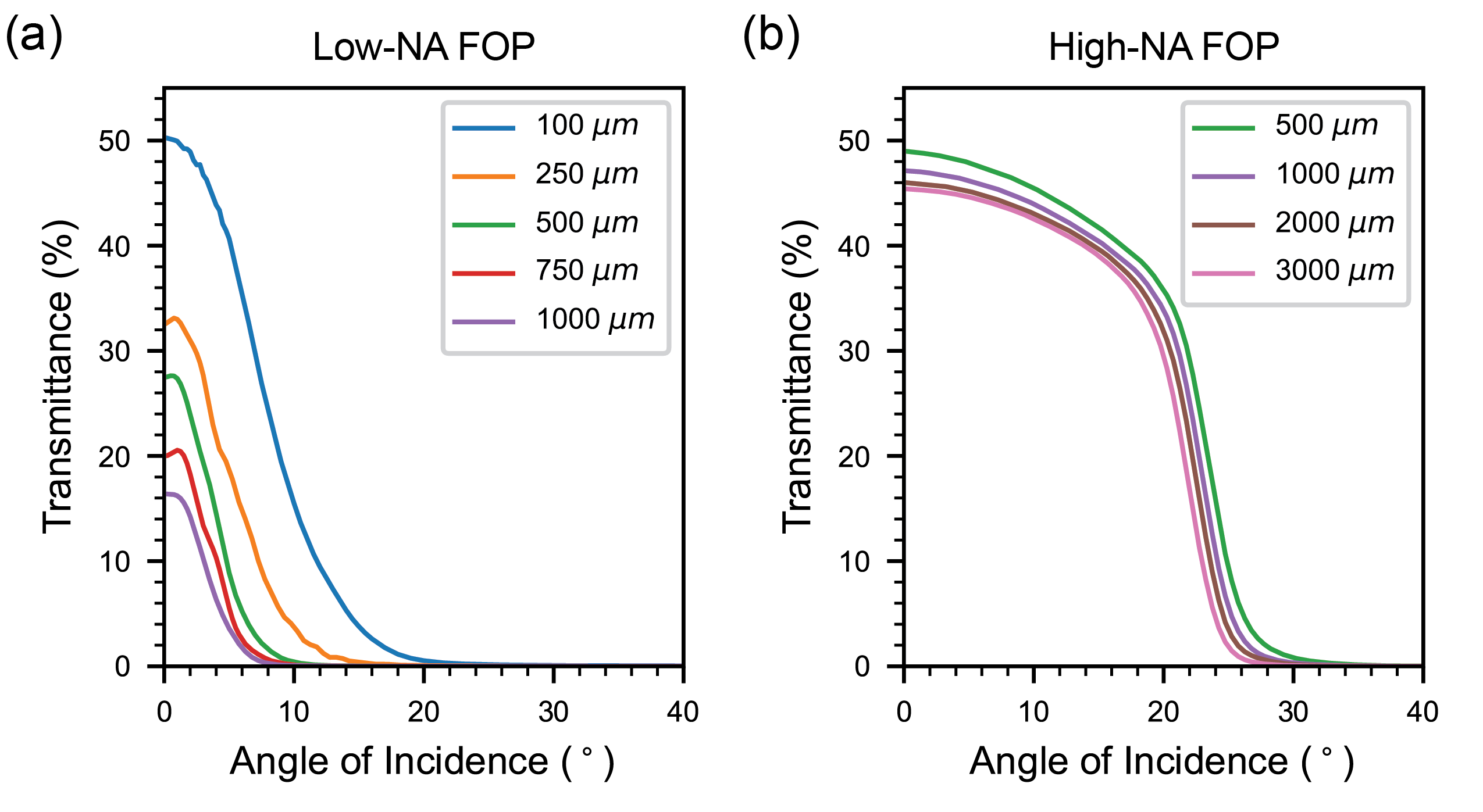}
\caption{\textbf{FOP characterization. }Measured transmittance across AOIs of a collimated 660 nm laser for different thicknesses of the (\textbf{a}) low-NA and (\textbf{b}) high-NA FOPs.}
\label{lin thickness sweeps}
\end{figure}

\begin{figure}[htbp]
\centering\includegraphics[]{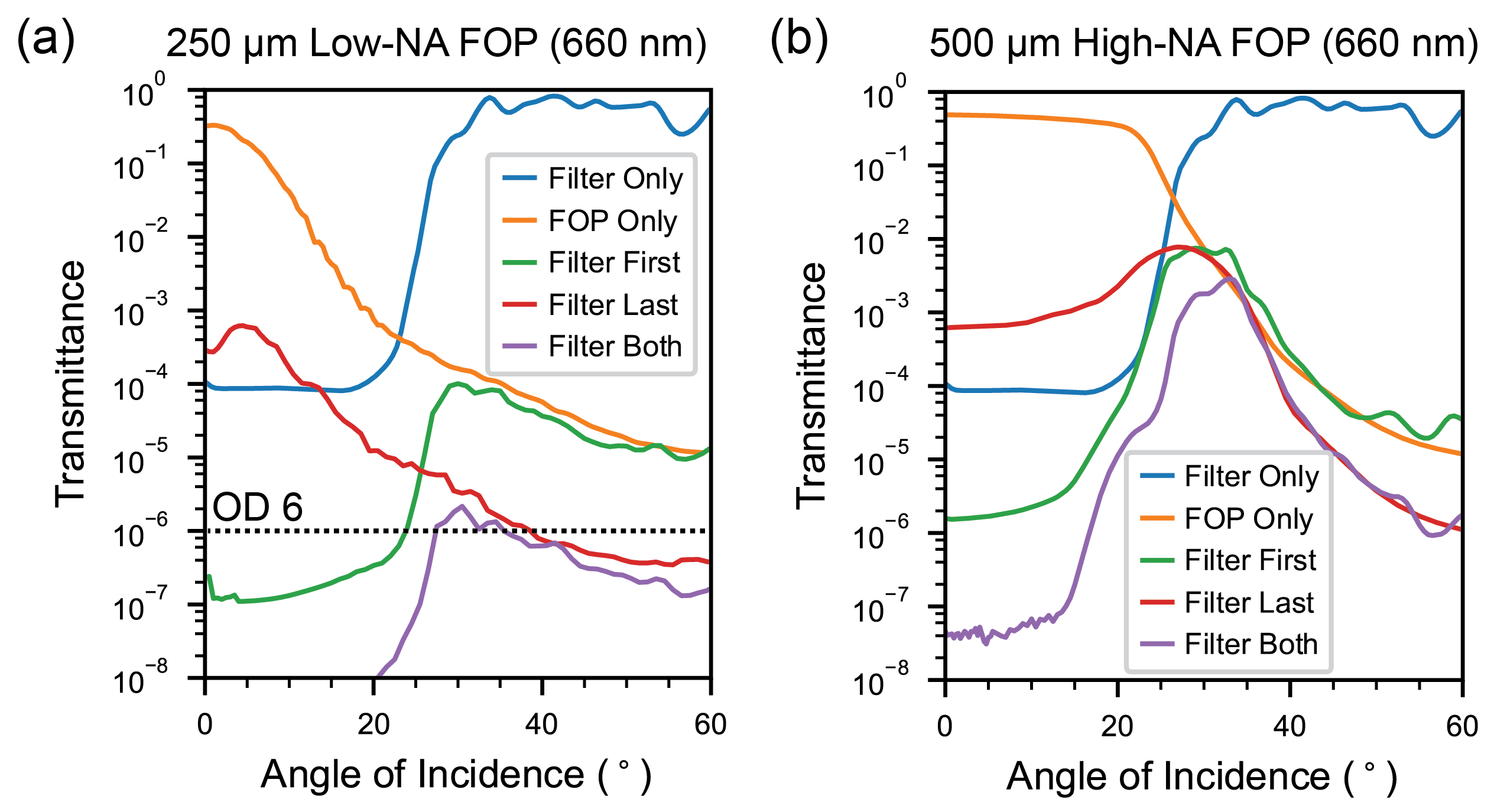}
\caption{\textbf{Optical frontend characterization. }(\textbf{a}) Measured transmittance across AOIs of a collimated 660 nm laser through the 250-$\mu$m-thick low-NA FOP, the interference filter, and different combinations of the filter and the FOP. (\textbf{b}) The same measurement for the 500-$\mu$m-thick high-NA FOP. }
\label{filter configurations}
\end{figure}

\begin{figure}[htbp]
\centering\includegraphics[]{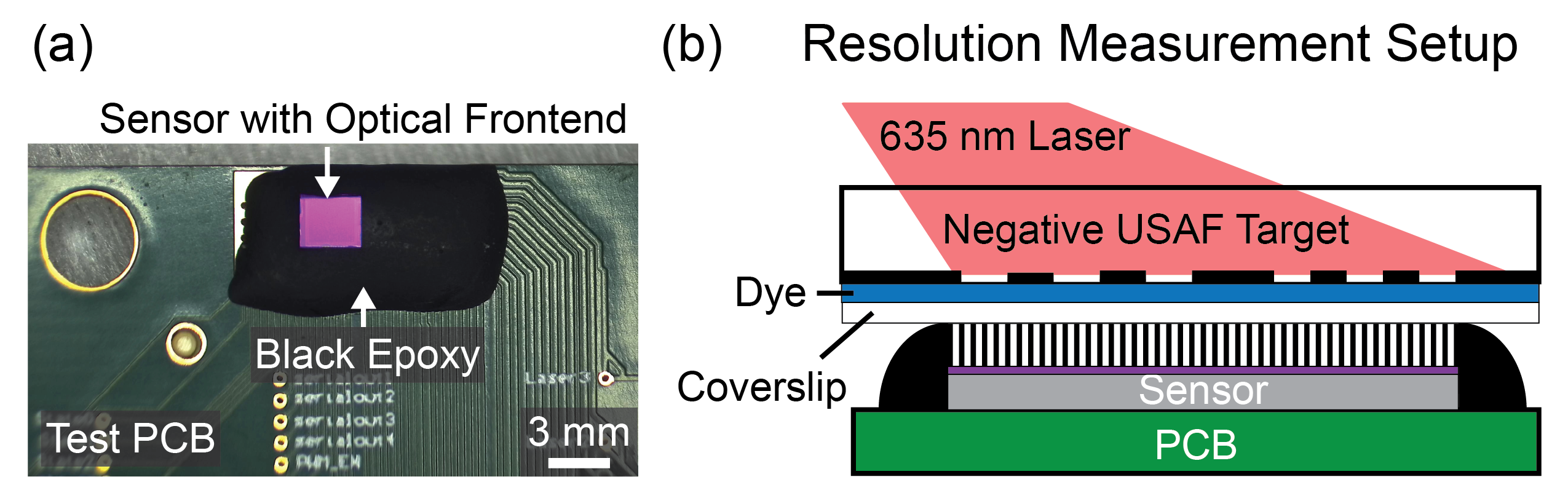}
\caption{\textbf{Imaging experiment setups.} (\textbf{a}) Photo of the sensor with the optical frontend. (\textbf{b}) Diagram of the setup for resolution measurements.}
\label{imaging setup}
\end{figure}

\begin{table}[htbp]
\caption{Comparison of recently published filters for lensless fluorescence sensors.}
\centering\includegraphics[]{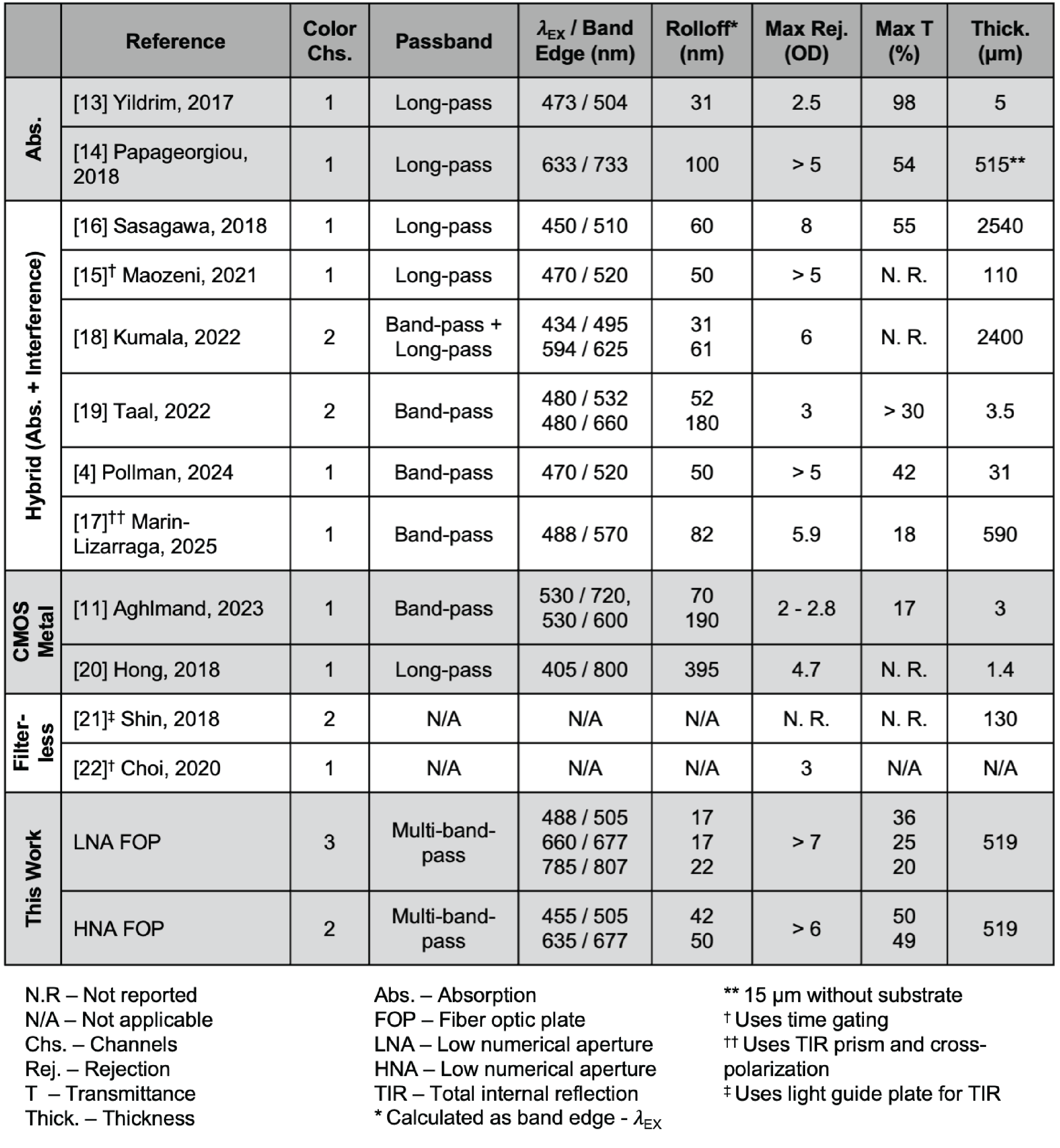}
\label{comparison table}
\end{table}

\end{document}